\documentclass[twocolumn,showpacs,amsmath,amssymb,aps,pre,reprint,longbibliography]{revtex4-1}
\usepackage{graphicx}
\usepackage{xcolor}
\usepackage[hidelinks]{hyperref}
\begin{document}

\title{
Pattern Formation and Flocking for Particles Near The Jamming Transition on
Resource Gradient Substrates} 
\author{
L. Varga$^{1}$, A. Lib{\' a}l$^{1}$, C. Reichhardt$^{2}$, and C. J. O. Reichhardt$^{2}$ 
} 
\affiliation{
$^{1}$ Mathematics and Computer Science Department, Babe{\c s}-Bolyai University, Cluj-Napoca 400084, Romania\\
$^{2}$ Theoretical Division and Center for Nonlinear Studies,
Los Alamos National Laboratory, Los Alamos, New Mexico 87545, USA}

\date{\today}
\begin{abstract}
We numerically
examine a bidisperse system of active and passive particles coupled to 
a resource substrate. The active particles deplete the resource at a fixed
rate and move toward regions with higher resources,
while
all of the particles interact sterically with each other.
We show that at high densities,
this system exhibits a rich variety 
of pattern forming phases
along with directed motion or flocking
as a function of the relative rates of resource absorption
and consumption as well as the active to passive particle ratio.
These include partial phase separation
into rivers of active particles flowing through passive clusters,
strongly phase separated states where the active particles induce 
crystallization of the passive particles,
mixed jammed states, and fluctuating mixed fluid phases. 
For higher resource recovery rates, we
demonstrate that the active particles can undergo
motility induced phase separation,
while at high densities, there can be
a coherent flock containing only active
particles or
a solid mixture of active and passive particles.
The directed flocking motion typically shows a transient
in which the flow switches among different directions 
before settling into one direction,
and there is a critical density below which
flocking does not occur. We map out the different phases 
as function of system density, resource absorption and recovery rates,
and the ratio of active to passive particles. 
\end{abstract}
\maketitle
  
\section{Introduction}

There is a large class of particle-like
systems that can exhibit collective motion when coupled to a substrate.
In some cases, the motion
is the result of driving by an external bias,
such as vortices
in type-II superconductors driven
over random landscapes \cite{Bhattacharya93,Blatter94},
moving charged colloidal particles interacting with
a rough landscape \cite{Reichhardt02,Pertsinidis08,Tierno12a},
frictional systems \cite{Vanossi13}, and
the larger class of systems
that exhibit collective depinning phenomena \cite{Fisher98,Reichhardt17}.
As a function of
increasing external bias or changing particle density,
these systems exhibit
distinct types of flow patterns.
In elastic flow, all of the particles travel at the same speed
and keep the same neighbors,
while in plastic flow, the motion
is broken 
into rivers or channels of flow
where portions of the system are pinned while other
portions are mobile \cite{Fisher98,Reichhardt17}.
Often,
transitions appear between different flow states as a function of driving,
such as the dynamical ordering of a
plastically moving phase into a moving crystal or moving smectic phase
\cite{Pardo98,Olson98a,Reichhardt17}.
In bidisperse particle assemblies
where
each
particle species
has a different coupling to
the external biasing or the substrate,
mixed flow
or fully phase separated states can appear.
An example of this
is the laning transition studied for bidisperse
particles
moving in
opposite directions 
\cite{Dzubiella02a,Vissers11a}, where
possible dynamic phases include
spatially separated lanes
as well as
mixed flowing or jammed configurations \cite{Reichhardt18}.

Active matter or self-propelled particles represent
another system that can exhibit
collective dynamic phases
\cite{Marchetti13,Bechinger16},
such as fluid states,
directed motion, and motility
induced phase separated states in which an active solid coexists with a low
density active fluid
\cite{Fily12,Redner13,Palacci13,Buttinoni13,Cates15,Reichhardt15}.
In the presence of a substrate,
additional phases appear,
including
aligned lanes or plastically
moving states with
coexisting regions of moving and immobile particles \cite{Sandor17a}.
In flocking systems represented by the Vicsek model
\cite{Vicsek12,Morin17},
the particles form flocks that move in a fixed direction at constant velocity.
In many
flocking systems,
there is a critical density
at which a transition occurs from a fluid
state into a coherently moving state.

More recently, another type of active matter system was introduced in 
which the motion of the particles is
produced via feedback from a resource substrate \cite{Wang21}.
This model was initially proposed for
robot swarms on resource landscapes
that are intended to mimic 
the motions of foragers.
The robots or moving particles
deplete the resources at a certain rate
and the resources recover at a different rate.
The particles are subjected to a force
termed a field drive that
moves them toward regions
with larger resource concentrations
\cite{Wang21}.
This system shows a series of
dynamic phases including liquid, crystalline, and jammed states
as the resource consumption and replenishment rates are varied.
A modification of this model
was used to study 
the addition of evolutionary rules to the robots \cite{Wang22}.
A different variation of this model
employed particles coupled to an
array of substrate sites that each have a resource parameter \cite{Varga22}.
Any particle that covers a site absorbs the resources at
a rate $r_{\rm abs}$, while at the same time,
the sites recover the
resources at another rate $r_{\rm rec}$ up to
a maximum resource
level.
A two-dimensional realization of this model was explored
for interacting disks
at an area coverage or density
of $\phi = 0.549$, well below the
jamming density of $\phi=0.9$
expected for monodisperse disks \cite{Reichhardt14}.
At these lower densities,
distinct phases appear, including a liquid-like continuously
fluctuating phase when the resource absorption and recovery
rates are similar
and
a frozen phase when the recovery rate is so high
that the
resource gradient vanishes and the particles experience no
driving force.
At low recovery rates, pulsating motion arises that is
similar
to the patterns found in
excitable media \cite{Zykov18}.
The pulsating waves form
when the resources must build up over time,
so that when the resource gradient becomes large enough, the particles
move in a front.
This causes the resource levels to
drop, and the particles remain frozen
again
until the resources are sufficiently replenished
to generate
the next wave.
The effect of adding passive particles, which interact only with
other particles but not with the substrate, was also considered.
For high resource recovery rates, the active particles
can push 
the passive particles
together
into a high density crystalline
passive phase surrounded by a
lower density fluid of active particles and a small number of passive
particles. This phase separation resembles
the motility
induced phase separation observed in active matter systems \cite{Fily12,Redner13,Palacci13,Buttinoni13,Reichhardt15,Cates15},
except it occurs for the passive particles.

In this work we study the high density limit of the
resource landscape model introduced in Ref.~\cite{Varga22},
and consider the jamming of
active and passive particle mixtures
as well as systems containing only active particles.
We concentrate on
three regimes.
In the scarce regime, the
resource absorption rate is high and the recovery rate is low; in 
the balanced regime, the absorption rate is about
four times higher than the recovery rate;
and in the plentiful regime, the recovery rate and
absorption rates are nearly equal.
In the scarce regime,
we find two types of
phase separated states including a partial clustering regime
and a river-like regime
where the active particles move in winding paths between 
islands of passive particles.
At the highest densities,
a mixed jammed state can form.
In the balanced regime,
we find strong phase separation
when the active particles push
the passive particles into dense crystalline arrangements with
cluster, stripe, void and uniform morphologies
similar to those found
in equilibrium phase separating systems
\cite{Seul95,Malescio03,Reichhardt10,Neto22}.
The active particles
themselves can also form crystalline motility induced
phase separated clusters where the activity induces the self-clustering.
At high densities, the
clusters undergo switching directed motion in which
flow persists in one direction for a period of time before switching
to a new direction,
and at even higher densities,
a flocking state appears
where clusters containing a mixture of active and passive
particles continuously move
in a fixed direction, similar to
what is observed in the Vicsek model
\cite{Vicsek12,Morin17}.
Within the balanced regime, flocking can only occur in the presence
of passive particles,
which are necessary to produce asymmetric resource gradients in the substrate.
In the plentiful regime,
flocking is possible even if all of the particles are active, and
there is also a phase
where the active particles flock while the passive particles form
a stationary crystalline solid.
We find that the velocity of the directed motion is non-monotonic
as a function of recovery rate, starting at a low value
for low recovery rates 
and increasing with increasing recovery rate,
but dropping back to a value close to zero
at higher recovery rates
when the
resource gradients become very weak or are absent.

\section{Simulation}

\begin{figure}
\includegraphics[width=\columnwidth]{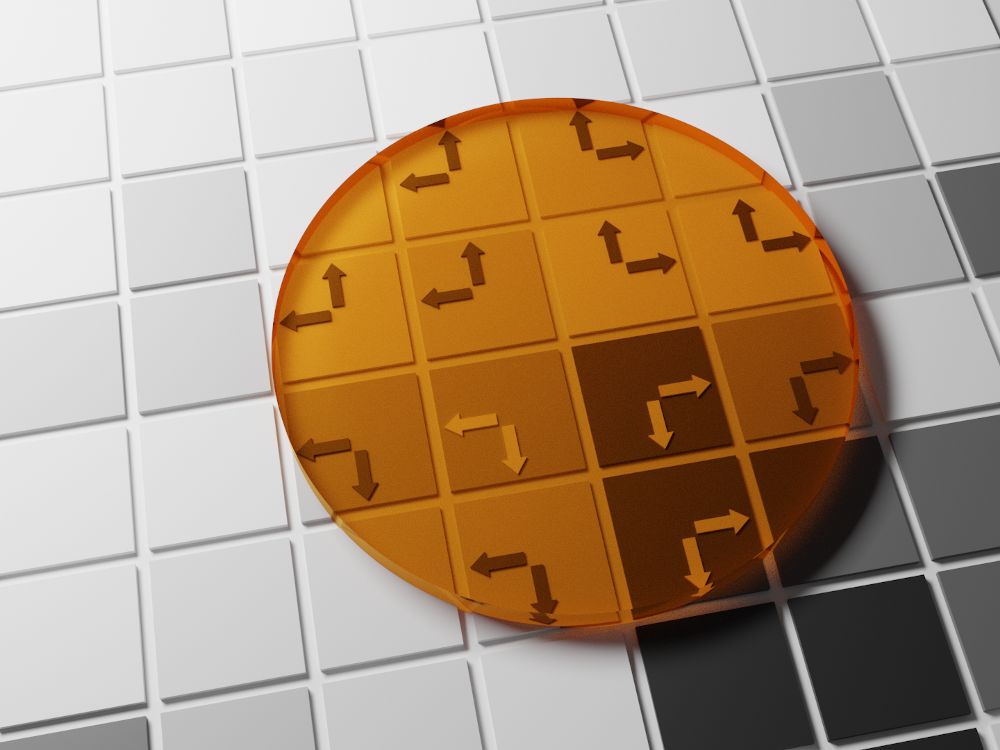}
\caption{Schematic illustration of the interaction of
an active particle (orange disk) with the resource substrate (boxes).
The resource sites may have high (white) or low (black) levels of
resource $S_g$ at any given time step.
Each particle occludes $12$ grid sites in a cross
geometry, indicated by the boxes containing arrows.
To represent a resource gradient-induced drive,
an occluded grid site $i$ exerts
a force on the particle
in the direction of each arrow with a magnitude proportional
to $S_g^i$.
If the resource concentration
is flat in any direction, the particle will experience
no net force in that direction.
The active particle depletes the resource in each of the 12 grid
sites at a rate $r_{\rm abs}$, while all of the grid sites recover resources
at a rate $r_{\rm rec}$ up to a maximum level of $S_g^i=1$.
}
\label{fig:1}
\end{figure}

We simulate a system containing an active substrate of size $L_x \times L_y = 100\times 200$ 
that is covered by a grid of $200\times 400$ square cells of size
$l_g \times l_g$, where $l_g=0.5$. Each grid cell $i$ has a 
resource value $S_g^i$ that can change continuously between $S_g^i=0$ and
$S_g^i=1$.
An assembly of $N$ disk-like particles with radius $R=1.0$ is placed on
the sample, which has periodic boundary conditions in the $x$ and $y$
directions.
Particles denoted as active
couple to the substrate and interact sterically with other
particles, while passive
particles do not couple to the substrate and
experience only steric particle-particle interactions.
During each time step,
there is a recovery and absorption
stage and a movement stage, in that order.

In the recovery and absorption stage, 
each active particle interacts with the $12$ cells closest to it on the
underlying grid, which form a $4 \times 4$ cross-like pattern
with the corner cells missing, as illustrated in Fig.~\ref{fig:1}. The
center of the particle is above one of the four central grid cells
in this pattern.
The amount of resource $S_g^i$ at cell $i$ is
first subjected to recovery, and is
then absorbed if the cell is occluded by an active particle.
The overall equation describing the resource level evolution is
$S_g^i(t+\Delta t)=S_g^i(t) + r_{\rm rec} -O_g^i r_{\rm abs}$
where $S_g^i(t)$ is the amount of resource in cell $i$ at the beginning
of the recovery stage,
$r_{\rm rec}$ is the resource recovery rate,
$r_{\rm abs}$ is the resource absorption rate by the active particle,
$O_g^i = 1 (0)$ if the grid site is occluded (unoccupied), and
a factor of the simulation time step 
$\Delta t=0.005$ has been folded in to the definitions of
$r_{\rm rec}$ and $r_{\rm abs}$.
Due to the fact that the value of $S_g^i$ is bounded within the range
$0 \leq S_g^i \leq 1$, in our actual implementation the recovery of
the resource cells is computed separately from the absorption,
with recovery calculated first followed by calculation of absorption.
Recovery occurs regardless of whether the cell is occluded by an
active particle, and it is subject to the constraint that the resource
level of any cell cannot go above $S_g^i=1$.
Cells that are occluded by an active particle then experience an
absorption
step subject to the constraint that the amount of
resource $S_g^i$ at a cell is not allowed to drop below $S_g^i=0$.
Cells that are not occluded by an active particle experience no
absorption.

In the movement stage, particle-particle interaction forces are computed and
added to the driving force from the grid on the active particles, giving
an overdamped equation of motion of
\begin{equation}
  \eta {\bf v}_i = {\bf F}^g_i + {\bf F}^{pp}_i
\end{equation}
for particle $i$, where ${\bf v}_i=d{\bf r}_i/dt$ is the velocity of the
particle and $\eta=1$ is the damping constant.
An occluded cell $k$ exerts a force ${\bf f}_k^x + {\bf f}_k^y$
in both the $x$ and $y$ directions on the center of the active particle that
is proportional to the resource level $S_g^k(t)$ at that cell
and independent of the distance from the cell
center to the particle center. 
The sign of each force component
is determined by the position of the center of the cell
relative to the particle center, and is positive for cells
whose centers are above or
to the right of the particle center, and negative for cells
whose centers are below or to the left
of the particle center.
The total driving force on the particle is
${\bf F}^g=\sum_{k=1}^{12}({\bf f}_k^x + {\bf f}_k^y)$.
If the resource
concentration gradient is flat along one direction, the forces
exerted by the cells cancel in that direction.
The theoretical maximum value of ${\bf F}_i^g$ for a situation in
which all of the cells on one side of the particle center have
a full resource level of $S_g^k=1$ and the remaining cells on
the other side of the particle center are entirely depleted with
$S_g^k=0$ is $|{\bf F}_i^g|=6$.
For passive particles, ${\bf F}_i^g=0$ regardless of the amount of
resource present in the grid sites.
Particle-particle interactions
are given by a steric harmonic repulsive force,
${\bf F}^{pp}=\sum_{j=1}^{N_p} k(d-r_{ij})\Theta(d-r_{ij}){\bf \hat r}_{ij}$.
Here $r_{ij}=|{\bf r}_i-{\bf r}_j|$,
${\bf \hat r}_{ij}=({\bf r}_i-{\bf r}_j)/r_{ij}$,
the spring constant is $k=20.0$,
$\Theta$ is the  Heaviside step function, and
$d=2R$ where $R$ is the particle radius.
Integration of the equations of motion
is performed with a velocity Verlet routine.

In this work we consider total particle densities
$\phi=N\pi R^2/(L_x L_y)$
ranging from $\phi=0.75$ up to $\phi=0.935$.
For a monodisperse disk packing, the jamming density
where the system forms 
a triangular lattice
is just above $\phi = 0.91$
\cite{Reichhardt14}.
Thus, for densities above $\phi=0.91$ we pass from the hard disk
limit to the foam limit.
The system contains $N_{p}$ passive particles and $N_{a}$ active particles,
where $N_p+N_a=N$.
The passive particles interact sterically with the active particles but
have no interaction with the resource substrate.

\begin{figure}
\includegraphics[width=\columnwidth]{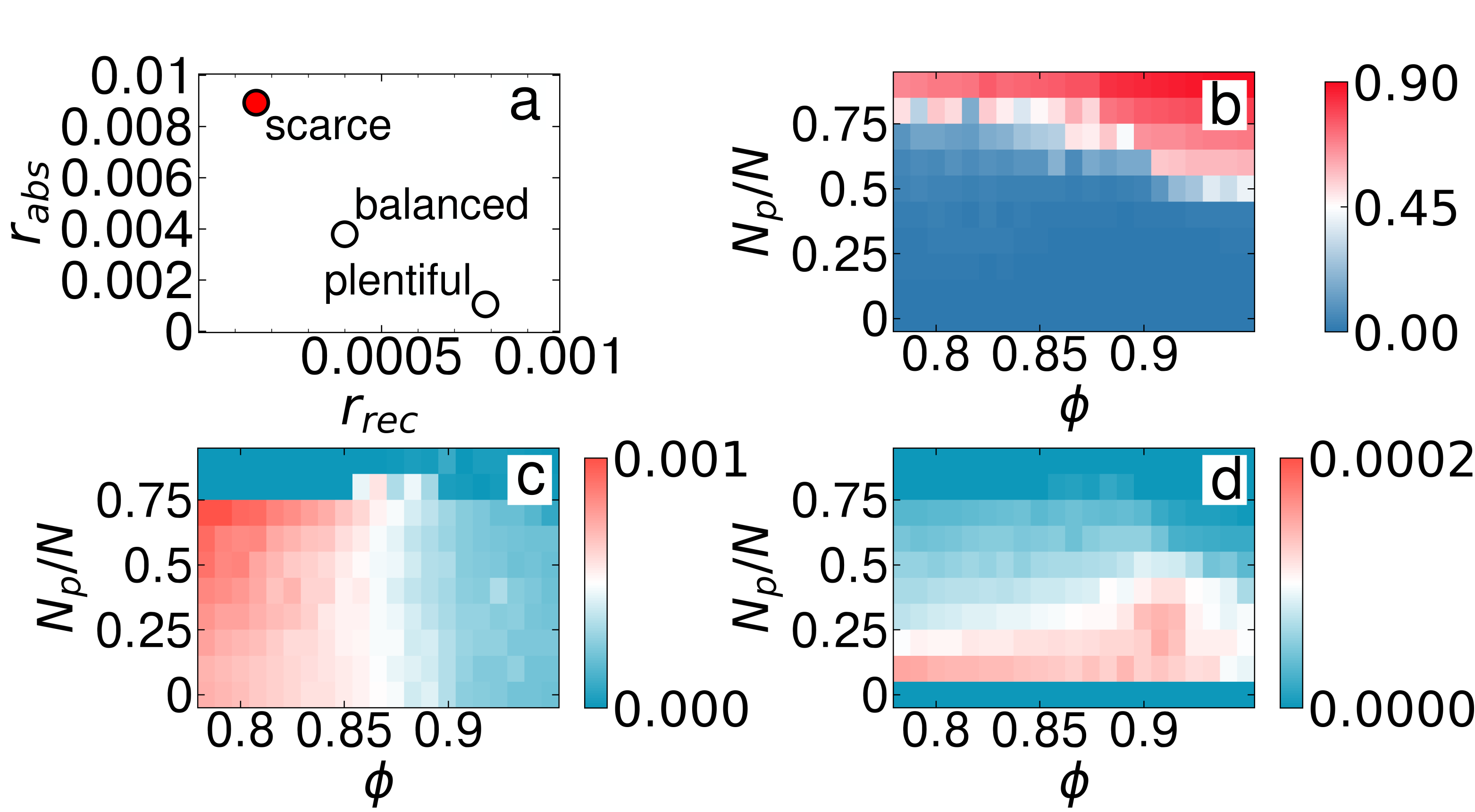}
\caption{(a) Illustration of the three regimes we consider as a function of
absorption rate $r_{\rm abs}$ versus recovery rate $r_{\rm rec}$:
scarce with $r_{\rm abs} = 0.0087$ and $r_{\rm rec} = 0.000175$;
balanced with $r_{\rm abs} = 0.0036$ and $r_{\rm rec} = 0.000425$;
and plentiful with $r_{\rm abs} = 0.0009$ and $r_{\rm rec} = 0.000825$.
(b-d) Results as a function of the fraction of passive particles
$N_p/N$ versus total density $\phi$ from a system in the scarce regime,
marked by a red dot in panel (a).
(b) Heat map of $C_L$, the fraction of particles
in the largest cluster.
(c) Heat map of $\langle |V_a|\rangle$, the average absolute value of the active particle velocities.
(d) Heat map of $\langle |V_b|\rangle$, the average absolute value of the passive particle velocities.
}
\label{fig:2}
\end{figure}

We focus on  three different regimes, shown in Fig.~\ref{fig:2}(a).
In the scarce regime, with
$r_{\rm abs} = 0.0087$ and $r_{\rm rec} = 0.000175$,
the absorption rate is high and the recovery rate is low.
In the balanced regime, with
$r_{\rm abs} = 0.0036$ and $r_{\rm rec} = 0.000425$),
the absorption rate is about four times higher than the recovery rate.
In the plentiful regime, with $r_{\rm abs} = 0.0009$
and $r_{\rm rec} = 0.000825$,
the absorption and recovery rates are nearly equal.
For each regime, we examine in detail the behavior of the
system as a function of two parameters: the density of particles and
the fraction of active particles.
We vary the overall particle number from $N = 4800$ to $6000$
giving a density range of $0.753 <  \phi < 0.942$, which spans the range
from below jamming to above jamming.
The fraction of passive particles $N_p/N$ varies from 0\% to 90\%.

As part of our analysis, we use
particle-particle contacts to define
clusters of particles that are touching one another but are
separated from other clusters or individual particles within the system.
To compute the fraction $C_L$ of particles in the largest cluster,
we identify the sizes $N_c$ of all of the clusters
in the system by counting
the number of particles in each cluster.
We then use the largest value of $N_c$ to define $C_L=N_c^{\rm max}/N$.
Additionally, we compute the
time average of the absolute value of the velocities
of the active,
$\langle |V_a|\rangle = \langle N_a^{-1}\sum_{i}^{N_a} |{\bf v}_i|\rangle$,
and passive,
$\langle |V_p|\rangle = \langle N_p^{-1}\sum_{i}^{N_p} |{\bf v}_i|\rangle$,
particles.

\section{Results}
\subsection{Scarce Regime}

We first focus on the scarce regime with
$r_{\rm abs} = 0.0087$ and $r_{\rm rec} = 0.000175$,
where 
$r_{\rm rec}/r_{\rm abs} = 0.02$ and the absorption rate is about 50 times
higher than the recovery rate.
In Fig.~\ref{fig:2}(b) we plot a heat map of $C_L$, the fraction
of particles that are in the largest cluster,
as a function of the fraction of passive particles $N_{p}/N$
versus the density $\phi$.
We find that strong clustering,
indicated by a large value of $C_L$, occurs
in the jammed regime at large $\phi$ when
the fraction of passive particles is greater than $N_p/N=0.5$.
Figure~\ref{fig:2}(c) shows a heat map
of $\langle |V_a|\rangle$, the average absolute value of the
active particle velocities,
as a function of $N_{p}/N$ versus $\phi$.
When the fraction of passive particles is large,
$\langle |V_a|\rangle$ is low,
but there is a local peak in $\langle |V_a|\rangle$ near
$N_p/N=0.7$. Below this peak, $\langle |V_a|\rangle$ diminishes
with decreasing $N_p/N$.
Additionally, for $\phi > 0.89$,
there is a drop in
$\langle |V_a|\rangle$ associated with the
onset of jamming or crystallization.
In Fig.~\ref{fig:2}(d) we plot a heat map of
$\langle |V_p|\rangle$, the
average absolute value of the passive particle velocities.
The largest $\langle |V_p|\rangle$
appears in a band around $N_p/N=0.1$ when there are
enough active particles to push the passive particles around,
but as $N_p/N$ increases, the fraction of active particles
that can perform the pushing decreases and
$\langle |V_p|\rangle$ drops.
There is also a drop in $\langle |V_p|\rangle$ at the highest
values of $\phi$ where
a jammed state
forms consisting of a mixture of active and passive particles
in a crystalline solid.

\begin{figure}
\includegraphics[width=\columnwidth]{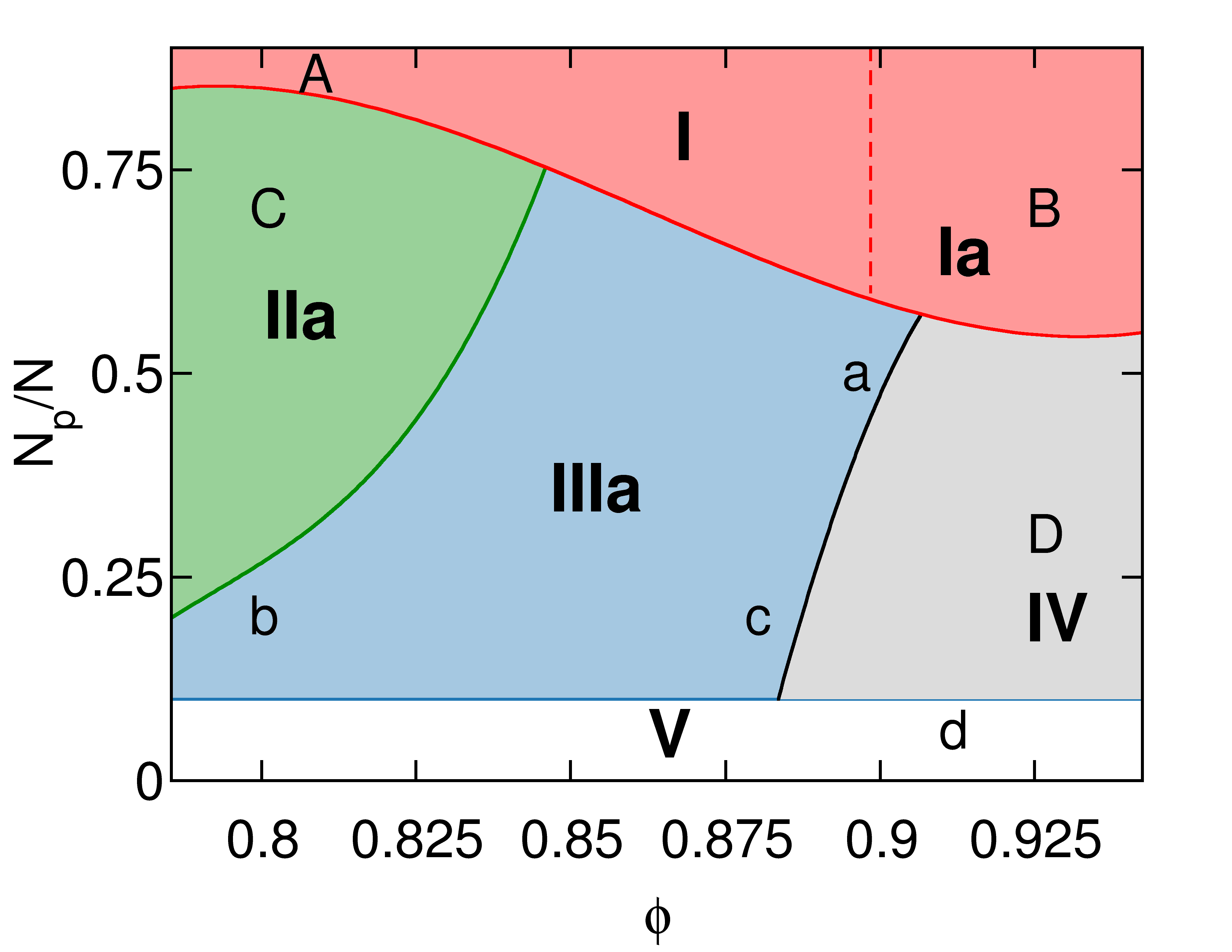}
\caption{Dynamic phase diagram as a function of $N_p/N$ versus $\phi$
for the scarce regime with $r_{\rm abs} = 0.0087$ and $r_{\rm rec} = 0.000175$.
The phases are:
I (large passive clusters, red below dashed line), I$_{a}$ (mixed jammed state,
red above dashed line), II$_{a}$ (rivers, green), III$_a$ (fluctuating, blue),
IV (random flocking, gray), and V (no passive particles, white).
Upper case labels indicate points at which the images in Fig.~\ref{fig:4}
were obtained, while lower case labels indicate points at which the
images in Fig.~\ref{fig:5} were obtained.
}
        \label{fig:3}
\end{figure}

From the quantities in Fig.~\ref{fig:2}  as well as images
of the system, we construct a dynamic phase diagram
for the scarce regime consisting of six
phases, shown in Fig.~\ref{fig:3} as a function of $N_p/N$ versus $\phi$.
In phase I, there is partial clustering of the passive particles
but the velocities of both passive and active particles are low.
Phase I$_a$,
which appears only when $N_{p}/N > 0.75$ and $\phi > 0.9$,
is a high density immobile jammed state in which
the active and passive particles are intermixed into a triangular
solid and the active particles are trapped by the surrounding
passive particles.
In phase II$_a$,
the passive particles form a series of high density solid islands
that exhibit almost no motion.
Riverlike structures
of active particles
move
between these islands
through confined winding channels that slowly change over time.
The channels are wide enough to permit the active
particles to form a low density liquid 
with higher velocities $\langle |V_a|\rangle$, as shown in Fig.~\ref{fig:2}(c).
In phase III$_a$,
there is weak clustering of the passive particles,
and the slowly moving active particles 
induce some motion of the passive particles.
Within phase IV, the passive and active particles move together 
in a particular direction for a period of time before switching to a
new collective direction of motion,
forming what we call a random flocking phase
in which the motion is coherent over short times but not
over long times.
At $N_p/N=0$, when only active particles are present, we find phase V,
which at lower densities is a liquid
and
for $\phi > 0.9$ is
a crystalline solid
that
moves in a changing
direction over time.

\begin{figure}
\includegraphics[width=\columnwidth]{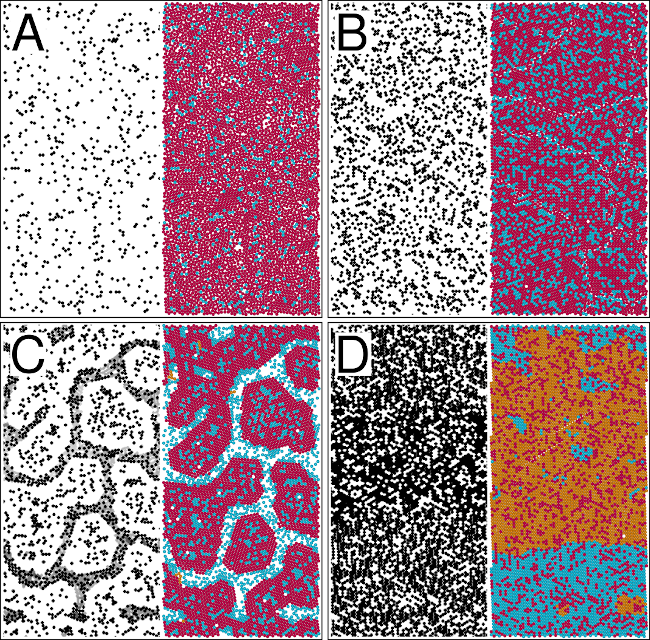}
\caption{
Simulation images from the scarce regime with $r_{\rm abs}=0.0087$
and $r_{\rm rec}=0.000175$. The left half of each panel shows a gray scale
map of the amount of resource present in the grid sites, with white indicating
$S^i_g=1$ or maximized and black indicating $S^i_g=0$ or fully depleted.
The right half of each panel shows the positions of the passive particles
(red), stationary or very slowly moving active particles (orange), and
active particles that are moving faster than a threshold velocity (blue).
Images correspond to the points marked A, B, C, and D in the phase diagram
of Fig.~\ref{fig:3}.
(A) Image from point A in phase I
with
$N_{p}/N = 0.9$ and
$\phi = 0.809$.
(B) Image from point B in phase I$_a$ or the jammed state
with $N_{p}/N = 0.927$ and $\phi = 0.7$.	
(C) Image from point C in phase II$_a$ at
$N_{p}/N = 0.7$ and $\phi = 0.801$,
showing riverlike ordering and flow.
(D) Image from point D in phase IV$_a$
at $N_{p}/N = 0.37$ and $\phi = 0.927$, 
where a coherently moving triangular lattice
switches between different directions of motion.
} 
\label{fig:4}
\end{figure}

In Fig.~\ref{fig:4} we show images from the simulations for the points
marked A, B, C, and D in the phase diagram of Fig.~\ref{fig:3}.
The left half of each panel shows the
state of the resource substrate on a
gray scale where darker color indicates greater depletion
of the resource site, while the right half of each panel indicates
the positions of the passive and active particles. The active particles are
colored according to their velocity,
where we use a cutoff velocity threshold of
$|{\bf v}_i|<1 \times 10^{-4}$
to indicate that an active particle is not moving or moving only
very slowly.
Figure~\ref{fig:4}(A)
illustrates phase I at
$N_{p}/N = 0.9$ and
$\phi = 0.809$,
where the velocities of both species are low.
Here,
the uniform background of passive particles is
modulated into local clusters of passive particles with triangular
ordering, and
the small number of individual active particles move slowly in the regions
between these clusters.
As $\phi$ increases, the clusters of passive particles percolate,
causing the cluster size $C_L$ found in Fig.~\ref{fig:2}(b) to increase.
This is correlated with
the appearance of
phase I$_a$,
illustrated in Fig.~\ref{fig:4}(B)
at $N_{p}/N = 0.7$ and $\phi = 0.927$.
Phase I$_a$ 
is 
a triangular jammed solid
composed of a mixture of active and passive particles,
where the triangular lattice contains a number of grain boundaries.
The active particles experience gradient-induced forces from the
substrate and try to move toward other sites,
but because the system is so dense, this motion is limited to the
production of local compression of the lattice,
which allows some lattice vacancies to form. 

In Fig.~\ref{fig:4}(C)
we show
phase II$_a$ at
$N_{p}/N = 0.7$ and
$\phi = 0.801$,
where the active particle velocity reaches
a maximum 
but the passive particle velocity
is zero, as indicated in
Fig.~\ref{fig:2}(c,d).
Here the
system forms a river-like state
where the active particles partially segregate into
low density regions of river-like paths.
Along these paths, the higher
concentration of active particles depletes the resource grid.
The passive particles are pushed together into dense islands
with local triangular ordering, but 
there are still some active particles trapped within these dense regions.
Since the passive particles
form isolated island-like structures that do not percolate,
$C_L$ does not become large in this regime.
Figure~\ref{fig:4}(D) illustrates 
phase IV
at $N_{p}/N = 0.37$ and $\phi = 0.927$,
where we find a uniform triangular lattice in which
both species are mixed. 
Here there are enough active particles present to cause patches of the
lattice to move together in a fixed direction for a period of time,
after which the direction of motion changes or a new patch begins to
move.
Since the system is so dense, the
passive particles are trapped 
in the triangular lattice of the active particles and must move
with them,
producing
the finite velocities of both passive and active particles
found in Fig.~\ref{fig:2}(c,d).

\begin{figure}
\includegraphics[width=\columnwidth]{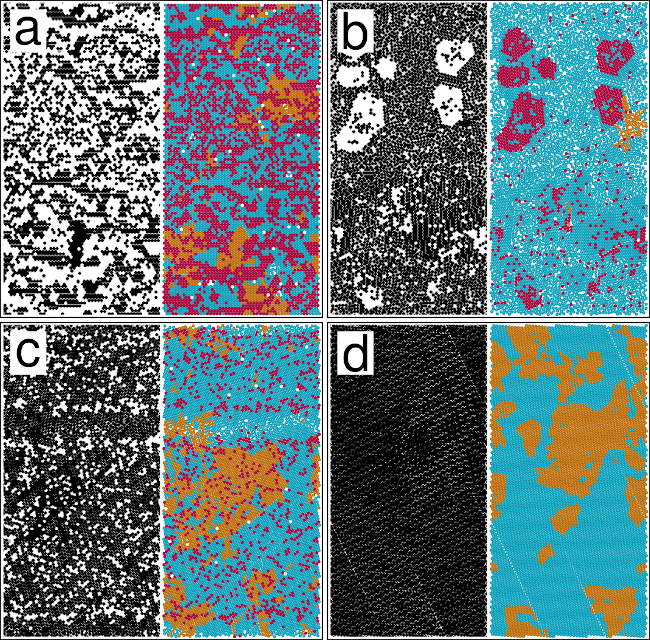}
\caption{
Simulation images from the scarce regime with $r_{\rm abs}=0.0087$
and $r_{\rm rec}=0.000175$. The left half of each panel shows a gray scale
map of the amount of resource present in the grid sites, with white indicating
$S^i_g=1$ or maximized and black indicating $S^i_g=0$ or fully depleted.
The right half of each panel shows the positions of the passive particles
(red), stationary or very slowly moving active particles (orange), and
active particles that are moving faster than a threshold velocity (blue).
Images correspond to the points marked a, b, c, and d in the phase diagram
of Fig.~\ref{fig:3}.
(a) Image from point a in phase III$_a$ at
$N_{p}/N = 0.5$ and
$\phi = 0.895$.
(b) Image from point b
in phase III$_a$ at
$N_p/N = 0.2$ and
$\phi = 0.801$.
(c) Image from point c
in phase III$_a$ at
$N_{p}/N = 0.2$ and $\phi = 0.88$ showing increased triangular ordering. 
(d) Image from point d
in phase V$_a$ at
$N_{p}/N = 0.0$ 
and $\phi = 0.91$,
where the system forms a triangular lattice of active particles.
}
\label{fig:5}
\end{figure}

In Fig.~\ref{fig:5}(a) we show the resource depletion
and the active and passive particle positions for phase
III$_{a}$ at $N_p/N=0.5$ and $\phi = 0.895$,
where there is a small amount of clustering of the
active particles leading to localized regions of motion.
Since the system is so dense, the motion takes the form of vacancy hopping
via the interchange of particles.
Phase III$_a$ is similar to the fluctuating  liquid 
phase found at lower densities in previous work \cite{Varga22}, but 
in this case there is no directed motion.  
Figure~\ref{fig:5}(b) shows
phase III$_a$
at $N_p/N = 0.2$  and
$\phi = 0.801$,
where
there is a stronger tendency for the
passive particles to phase separate from the active particles
to form local clusters.
There is, however, no large scale clustering of 
passive particles of the type found in phases I and II$_a$. 
In Fig.~\ref{fig:5}(c), phase III$_a$ at
$N_{p}/N = 0.2$ 
and $\phi = 0.88$
has large sections of triangular
ordering and the resources are more uniformly depleted.
At $N_{p}/N = 0.0$ and $\phi = 0.91$ in
phase V, shown in Fig.~\ref{fig:5}(d), the system is almost 
completely triangular. For lower densities, phase V becomes more fluid-like.  

\begin{figure}
\includegraphics[width=\columnwidth]{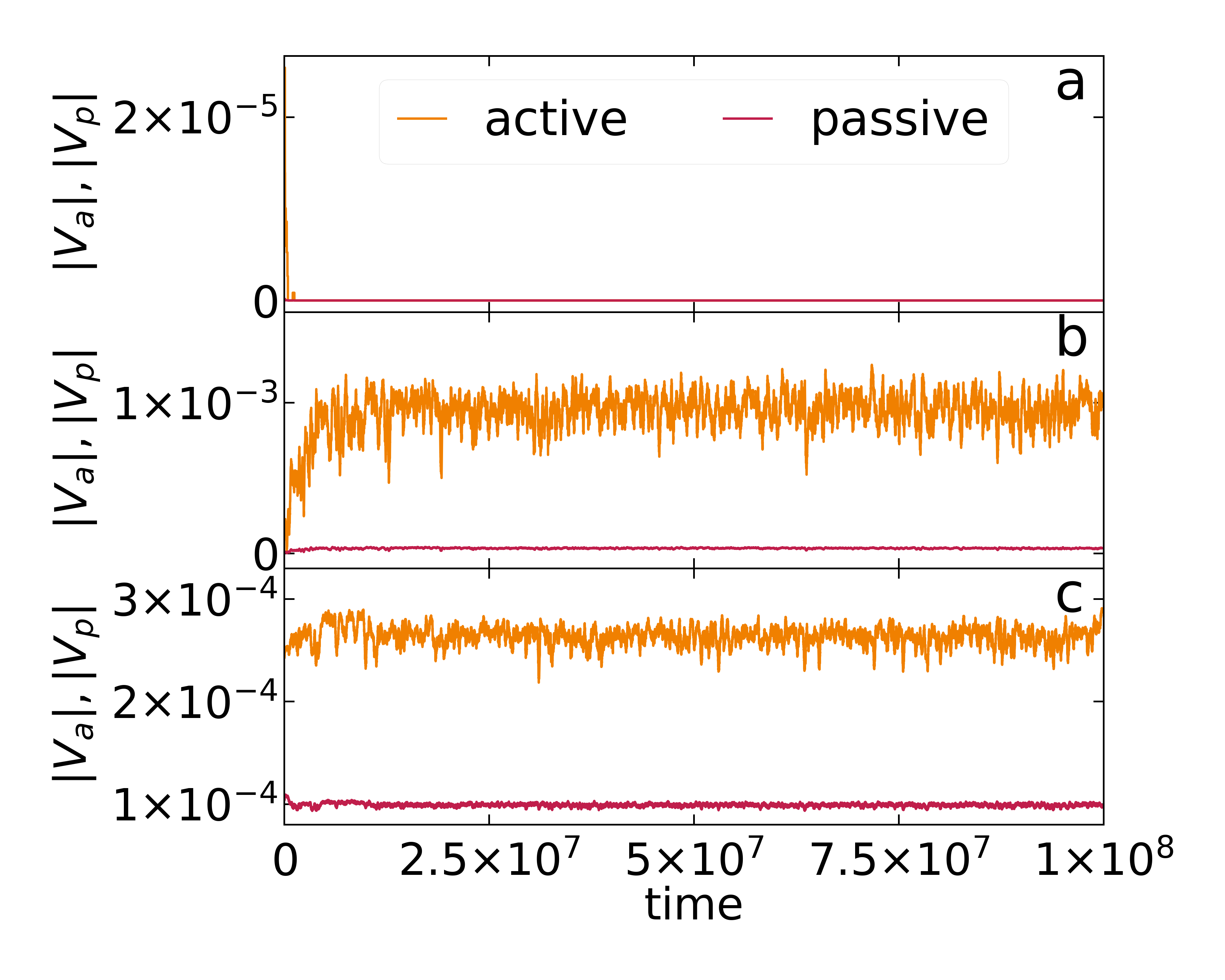}
\caption{
Time series of $|V_a|$ (orange) and $|V_p|$ (red)
in the scarce regime with $r_{\rm abs}=0.0087$ and $r_{\rm rec}=0.000175$
at labeled points from the phase diagram of Fig.~\ref{fig:3}.
(a) Phase I at point A with $N_p/N=0.9$ and $\phi=0.809$,
illustrated in Fig.~\ref{fig:4}(A),
showing no motion for either type of particle.
(b) Phase II$_a$ or the river-like flow phase at point C
with $N_p/N=0.7$ and $\phi=0.801$, illustrated in Fig.~\ref{fig:4}(C),
where
the active particles have a high velocity and the passive
particles have a low velocity.
(c) Phase IV at point D
with $N_p/N=0.37$ and $\phi=0.927$,
illustrated in Fig.~\ref{fig:4}(D),
showing a higher level of motion for both the active and the
passive particles.}
\label{fig:6}
\end{figure}

In Fig.~\ref{fig:6}(a)
we show the time series of
the absolute value of the active and passive
particle velocities, $|V_a|$ and $|V_p|$, for the scarce
regime with $r_{\rm abs}=0.0087$ and $r_{\rm rec}=0.000175$
in phase I from point A in the phase diagram of Fig.~\ref{fig:3}
at $N_p/N=0.9$ and $\phi=0.809$.
Here, the velocity of both types of particle is near zero.
For the phase II$_a$ river-like flow from point C in Fig.~\ref{fig:3}
at $N_p/N=0.7$ and $\phi=0.801$,
Fig.~\ref{fig:6}(b) shows that
the active particles have a finite velocity while the
velocity of the passive particles is near zero.
The plot also indicates that
there is a transient time during which
the active particles organize the passive particles
into islands, after which the system reaches a steady state. 
The fluctuations in $|V_p|$
in Fig.~\ref{fig:6}(b) are the result of collisions of active particles
with the edges of the passive particle islands.
In Fig.~\ref{fig:6}(c), the time series of $|V_a|$ and $|V_p|$
for phase IV from point D in Fig.~\ref{fig:3} at $N_p/N=0.37$ and
$\phi=0.927$ show that
both the active and the passive particles have finite velocities, 
but the active particles move about 2.5 times more rapidly than the
passive particles on average.

\subsection{Balanced Regime}

\begin{figure}
\includegraphics[width=\columnwidth]{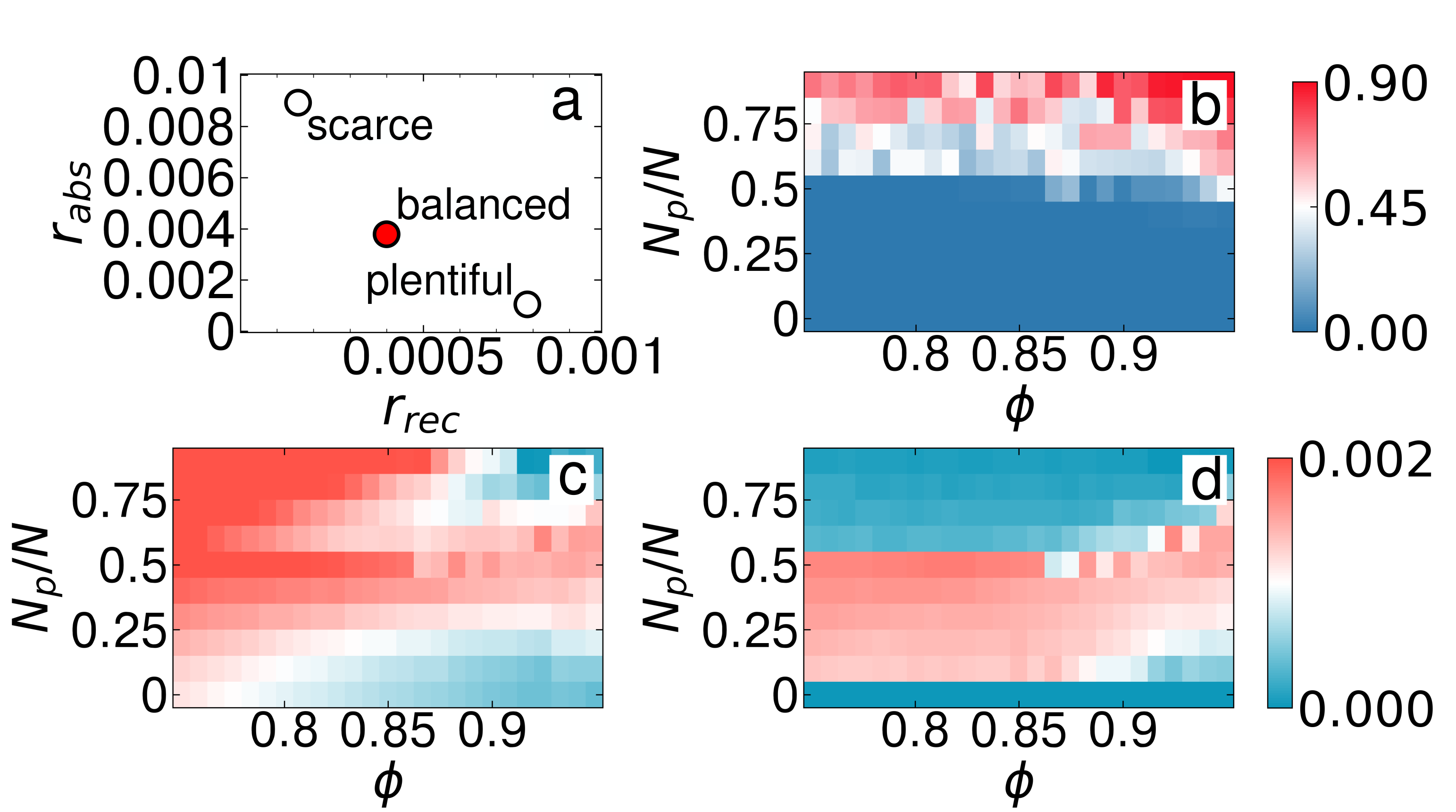}
        \caption{
(a) Diagram of the location of the scarce
($r_{\rm abs}=0.0087$, $r_{\rm rec}=0.000175$),
balanced
($r_{\rm abs}=0.0036$, $r_{\rm rec}=0.000425$),
and plentiful
($r_{\rm abs}=0.0009$, $r_{\rm rec}=0.000825$) regimes.
(b-d) Results as a function of the fraction of passive
particles $N_p/N$ versus total density $\phi$ from a system
in the balanced regime, marked by a red dot in panel (a), where
the ratio of absorption to recovery rates is $r_{\rm abs}/r_{\rm rec}=4.36$.
(b) Heat map of $C_L$, the fraction of particles in the largest cluster.
(c) Heat map of $\langle |V_a|\rangle$,
the average absolute value of the active particle velocities.
(d) Heat map of $\langle |V_b|\rangle$, the average absolute value
of the passive particle velocities.}
\label{fig:7}
\end{figure}

We next consider the balanced regime with $r_{\rm abs}=0.0036$
and $r_{\rm rec}=0.000825$, illustrated in Fig.~\ref{fig:7}(a), where 
the ratio of absorption rate to recovery rate
is around 4.36.
As a function of $N_p/N$ versus $\phi$, we plot
heat maps of the largest cluster size $C_L$ in
Fig.~\ref{fig:7}(b),
the average absolute active particle velocities
$\langle |V_a|\rangle$ in Fig.~\ref{fig:7}(c),
and the average absolute passive particle velocities
$\langle |V_p|\rangle$ in Fig.~\ref{fig:7}(d).
There is an extended
window in which
the velocity
of both the active and the passive particles is high.
For $0.754 < \phi < 0.87$ and $N_{p}/N > 0.4$, the
active particle velocity remains high
and does not drop even when  $N_{p}/N > 0.8$,
since in the fully phase separated system the active particles
can move more rapidly.

\begin{figure}
\includegraphics[width=\columnwidth]{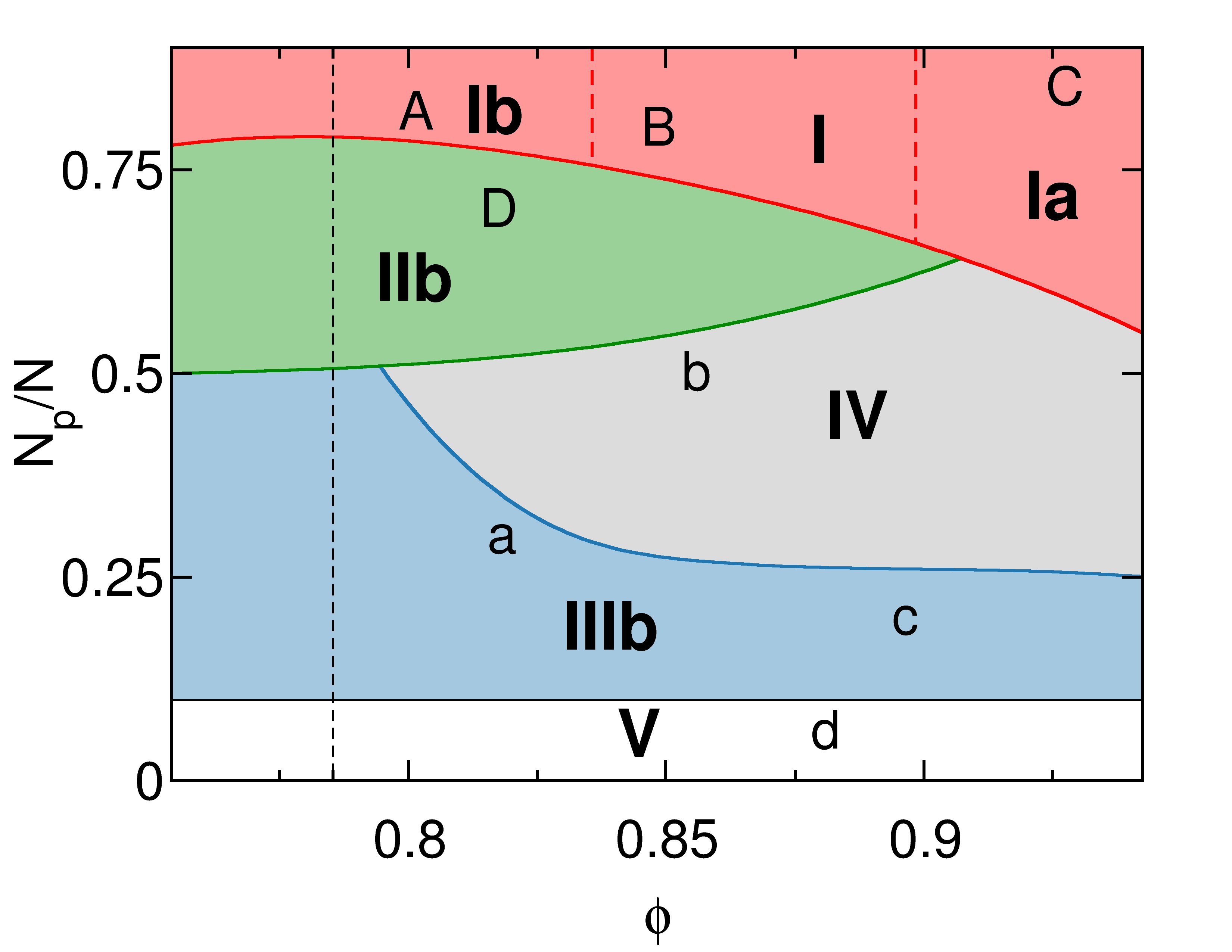}
\caption{Dynamic phase diagram as a function of $N_p/N$ versus $\phi$
for the balanced regime with $r_{\rm abs} = 0.0036$ and $r_{\rm rec} = 0.000425$. 
The phases are: I (phase separated cluster, red center),
I$_a$ (mixed jammed state, red right),
I$_b$ (stripe-like phase separation, red right),
II$_b$ (large rivers, green),
III$_b$ (active particle clustering, blue),
IV (directed motion, gray), and V (no passive particles, white).}
\label{fig:8}
\end{figure}

Using the features in Fig.~\ref{fig:7} along with the velocity
time series and images of the particle configurations,
in Fig.~\ref{fig:8}
we construct a dynamic phase diagram
as a function of $N_p/N$ versus $\phi$ highlighting the different phases.
In general, the balanced regime exhibits
extended regions of strong phase separation
in which the active particles can move rapidly  
and shepherd the passive particles into a cluster;
alternatively, the active particles themselves may undergo
motility induced phase separation.

\begin{figure}
\includegraphics[width=\columnwidth]{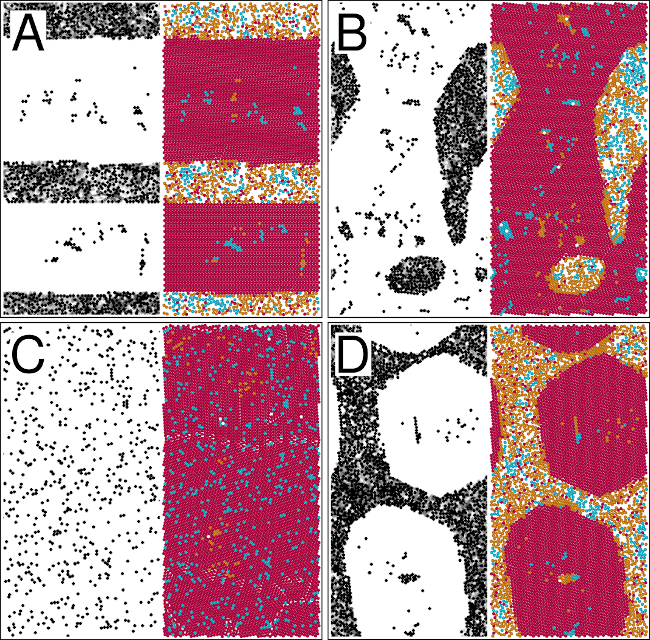}
\caption{Simulation images from the balanced regime with $r_{\rm abs}=0.0036$
and $r_{\rm rec}=0.000425$. The left half of each panel shows a gray scale
map of the amount of resource present in the grid sites, with white indicating
$S^i_g=1$ or maximized and black indicating $S^i_g=0$ or fully depleted.
The right half of each panel shows the positions of the passive particles
(red), stationary or very slowly moving active particles (orange), and
active particles that are moving faster than a threshold velocity (blue).
Images correspond to the points marked A, B, C, and D in the phase diagram
of Fig.~\ref{fig:8}.
(A) Image from point A in the low density phase I$_b$ at
$N_{p}/N=0.8$ and $\phi=0.801$.
(B) Image from point B in phase I
at $N_{p}/N=0.8$ and $\phi=0.848$ showing 
a phase separated void state.
(C) Image from point C 
in phase I$_a$ for $N_{p}/N=0.9$ and $\phi=0.927$  
showing a jammed state with triangular ordering.
(D) Image from point D in phase II$_{b}$ at
$N_p/N = 0.7$ and
$\phi = 0.817$.
}
\label{fig:9}
\end{figure}

For the balanced regime, the phase separation becomes much more pronounced in
phase I,
and the passive particles assemble into dense
clusters with triangular ordering.
This is illustrated in Fig.~\ref{fig:9}(A) for phase I$_b$
at $N_{p}/N=0.8$ and $\phi=0.801$, where the active particles
herd the passive particles into crystalline stripes.
In this case, the dense regions
have a local density just over $\phi_{\rm loc} = 0.91$,
while in the surrounding active particle regions,
the local density is well below $\phi_{\rm loc}=0.801$.
There is no directed motion of the active
particles,
and a small number
of active particles become trapped within the passive stripes.
From the heat map of $\langle |V_a|\rangle$ in Fig.~\ref{fig:7}(c),
we find that the absolute value of the active particle velocities
is actually higher for large $N_p/N$ than for $N_p/N=0$ where no
passive particles are present.
This seems counterintuitive since
it could be thought that passive particles can
only slow down the active particles; however,
when the passive particles phase separate into the stripes,
regions of no depletion appear underneath the stripes. As a result,
a large resource gradient forms at the edge of the stripes that
causes active particles along the stripe edges to move at high
velocities.
This can partially deform the surface of the passive stripe, and
indicates that
clusters or stripes of passive particles
can increase the active motion in the system
by
facilitating the creation of large
resource gradients.
Figure~\ref{fig:9}(B) shows that in phase I
at $N_p/N=0.8$ and $\phi=0.848$,
due to the higher density the passive particles can form a single large
connected cluster.
For even higher densities we find
phase I$_a$, as illustrated
in Fig.~\ref{fig:9}(C)
for $N_{p}/N=0.9$ and $\phi =  0.927$.
This is a uniform jammed state in which some active
particles are mixed in with the passive particles.
Since the strong phase separation is lost,
the active particles do not experience an acceleration effect and
both $\langle |V_a|\rangle$ and $\langle |V_b|\rangle$ are low.

Phase II$_a$ from the sparse regime is replaced
in the balanced regime by
phase II$_b$,
which is still a river-like phase but with much larger rivers that
form large scale phase separated structures,
as shown in Fig.~\ref{fig:9}(D) at
$N_p/N = 0.7$ and $\phi = 0.817$.
We note that the phase separated states
we observe in phases I, I$_a$, I$_c$, and II$_b$
have structures similar to those of phase separated states
that form in equilibrium pattern forming systems with
competing attractive and repulsive interactions.
Such systems exhibit bubble crystals, stripes,
void crystals, and uniform states
as a function of increasing density
\cite{Varga22,Seul95,Malescio03,Reichhardt10,Neto22}.
The formation of stripes or bubbles depends on the density.
In general, our system
forms
stripes or isolated bubbles
at lower passive particle densities,
void-like crystals
at intermediate densities,
and a uniform solid at high densities.
The clustering of the passive particles
due to the motion of active particles over a resource
substrate was also observed
in Ref.~\cite{Varga22} for much lower densities,
where only isolated islands formed.

\begin{figure}
\includegraphics[width=\columnwidth]{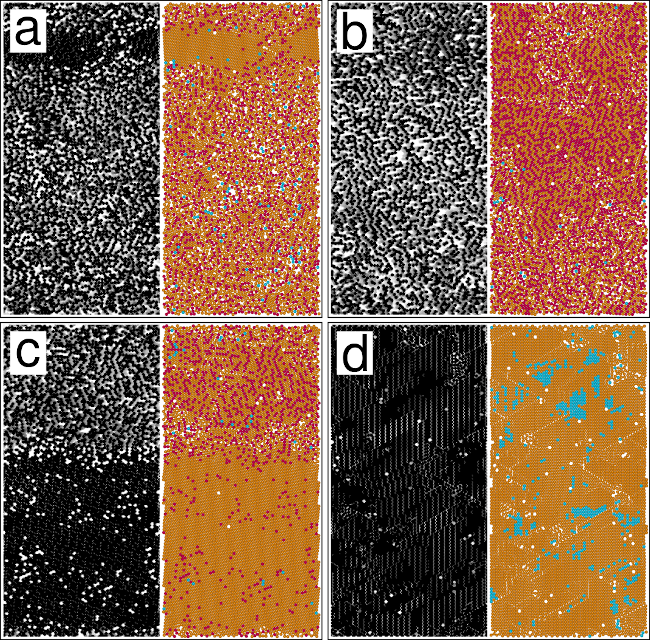}
\caption{Simulation images from the balanced regime with $r_{\rm abs}=0.0036$
and $r_{\rm rec}=0.000425$. The left half of each panel shows a gray scale
map of the amount of resource present in the grid sites, with white indicating
$S^i_g=1$ or maximized and black indicating $S^i_g=0$ or fully depleted.
The right half of each panel shows the positions of the passive particles
(red), stationary or very slowly moving active particles (orange), and
active particles that are moving faster than a threshold velocity (blue).
Images correspond to the points marked a, b, c, and d in the phase diagram
of Fig.~\ref{fig:8}.
(a) Image from point a in the
active clustering phase III$_b$
at $N_p/N=0.3$ and $\phi = 0.817$.
(b) Image from point b in phase IV
at 
$N_p/N = 0.5$ and
$\phi = 0.856$.
(c) Image from point c in phase III$_b$
at $N_p/N = 0.2$ and
$\phi = 0.895$.
(d) Image from point d
in phase V at $N_p/N=0$ and $\phi=0.88$.
}
\label{fig:10}
\end{figure}

Phase III$_b$ in Fig.~\ref{fig:8} is also phase
separated, but here the active particles form clusters while
the passive particles remain in a disordered lower density
liquid state,
as shown in Fig.~\ref{fig:10}(a)
at $N_p/N=0.3$ and $\phi=0.817$ and
in Fig.~\ref{fig:10}(c) at
$N_p/N=0.2$ and $\phi=0.895$.
This state is similar to the
motility induced phase separation
found in other active matter
systems
\cite{Fily12,Redner13,Palacci13,Buttinoni13,Reichhardt15,Cates15}, and
indicates that the motion of the active
particles is now large enough
to permit the active
particles to form self-clustering states.
In our system, phase III$_b$ emerges
when the ratio of active to passive particles is
below $N_p/N=0.5$
and when the recovery rate is high enough
for the active particles to remain moving continuously.
In Fig.~\ref{fig:10}(a)
at $N_p/N=0.3$ and $\phi = 0.817$,
an active particle cluster with triangular ordering appears
in the upper half the panel. Underneath this cluster,
the resources are more strongly depleted, as indicated by the
resource gray scale panel.
In Fig.~\ref{fig:10}(c) at
$N/N_{p} = 0.2$ and
$\phi = 0.895$,
the active cluster is much larger
and the region under the active cluster
shows very strong resource depletion.
The size of the active cluster grows with decreasing $N_p/N$
and increasing $\phi$.

\begin{figure}
\includegraphics[width=\columnwidth]{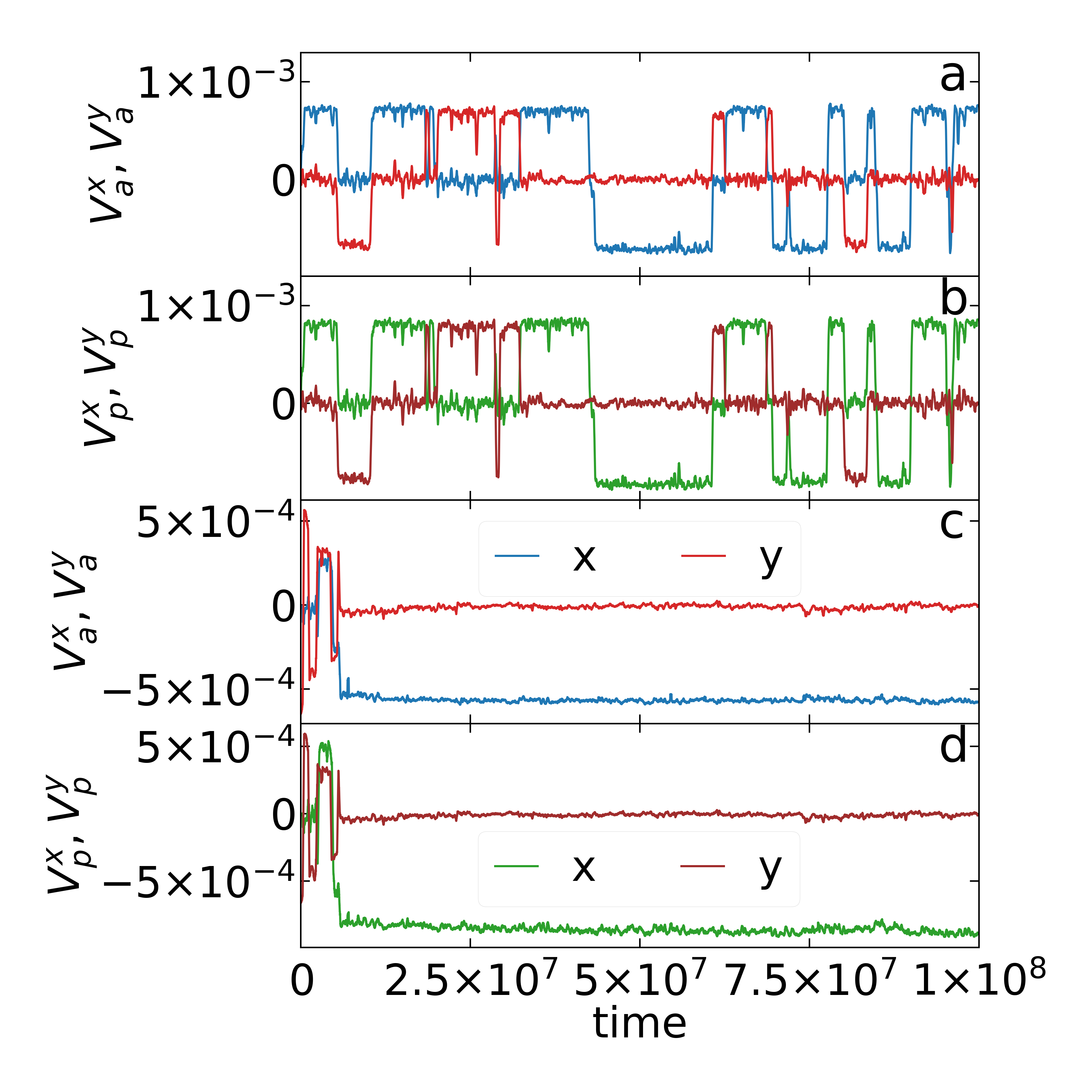}
\caption{
Time series of the active particle velocities in the
$x$, $V_a^x$ (blue), and $y$, $V_a^y$ (red), directions
and of the passive particle velocities in the
$x$, $V_p^x$ (green), and $y$, $V_p^y$ (maroon), directions
in the balanced
regime with $r_{\rm abs}=0.0036$ and $r_{\rm rec}=0.000425$ at labeled points
from the phase diagram of Fig.~\ref{fig:8}.
(a)  $V_a^x$ and $V_a^y$ for phase III$_b$ at point a with
$N_p/N=0.3$ and $\phi=0.817$,
illustrated in Fig.~\ref{fig:10}(a),
where there is switching directed motion.
(b)
The corresponding
$V_p^x$
and
$V_p^y$
also show
switching directed motion, indicating that the active particles have
entrained the passive particles.
(c)
$V_a^x$
and $V_a^y$
in phase III$_b$ from point c with
$N_p/N=0.2$ and $\phi=0.895$, illustrated
 in Fig.~\ref{fig:10}(c).
(d) The corresponding
$V_p^x$
and $V_p^y$.
Here, the system settles
into a state with directed motion of both
particle species in the negative $x$ direction.
}\label{fig:11}
\end{figure}

In phase III$_b$
we find partial directed motion or flocking where
clusters of active particles move collectively in
a particular direction either
for a period of time or, at higher densities, permanently.
To illustrate this, in Fig.~\ref{fig:11}(a)
we plot time series of the active particle velocities in the
$x$ and $y$ directions,
$V_a^x=N_a^{-1}\sum_{i}^{N_a}{\bf v}_i \cdot {\bf \hat x}$
and $V_a^y=N_a^{-1}\sum_{i}^{N_a}{\bf v}_i \cdot {\bf \hat y}$, respectively,
for
phase III$_b$ from Fig.~\ref{fig:8} at $N_p/N=0.3$ and $\phi=0.817$.
There are windows of time during which
the active particles move in a particular direction as a coherent
flock
before
the motion switches into a new direction.
In this case, particles in the less dense regions
move faster than the particles in the dense regions.
The dense region
resembles
what is found in traditional motility induced phase separation
since the particles
within this region are moving in different directions
and collide with one another,
leading to a reduced velocity.
In the less dense region, the
passive particles are entrained by the
active particles
and have nearly the same velocity as the active particles,
as shown in Fig.~\ref{fig:11}(b) where we plot
$V_p^x=N_p^{-1}\sum_{i}^{N_p}{\bf v}_i \cdot {\bf \hat x}$
and $V_p^y=N_p^{-1}\sum_{i}^{N_p}{\bf v}_i \cdot {\bf \hat y}$ for
the passive particles.
At higher densities,
there is a change in phase III$_b$ from
directed motion in different directions
to a flocking state in which all of the
particles flow permanently in a fixed direction,
as shown by the time series of $V_a^x$ and $V_a^y$ in Fig.~\ref{fig:11}(c)
and of $V_p^x$ and $V_p^y$ in Fig.~\ref{fig:11}(c) for a sample
with $N_p/N=0.2$ and $\phi=0.895$.
Here, both the passive and active particles move
in the negative $x$ direction after a short
transient time.

\begin{figure}
\includegraphics[width=\columnwidth]{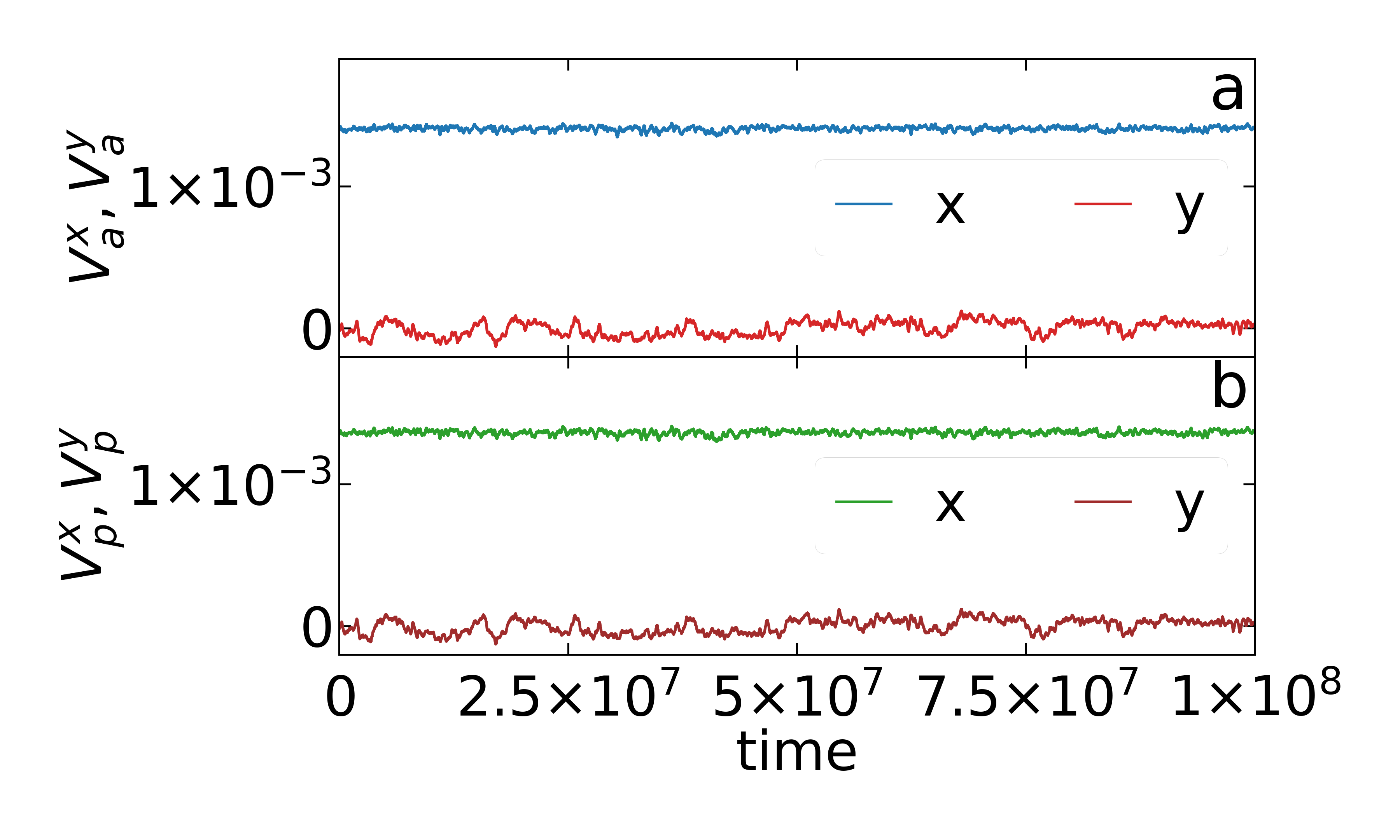}
\caption{
(a) Time series of $V_a^x$ (blue) and $V_a^y$ (red) for the active particles
in the balanced regime with $r_{\rm abs}=0.0036$ and $r_{\rm rec}=0.000425$
in phase IV
from point b of the phase diagram in Fig.~\ref{fig:8}
with
$N_p/N=0.5$ and $\phi=0.856$,  
illustrated in Fig.~\ref{fig:10}(b),
showing switching directed motion.
(b) The corresponding time series of $V_p^x$ (blue) and $V_p^y$ (maroon) for
the passive particles.
}
\label{fig:12}            
\end{figure}

Figure~\ref{fig:10}(b) illustrates
phase IV
at $N/N_{p} = 0.5$ and
$\phi = 0.856$,
where the active and passive particles form a mixed cluster that
depletes the resources in a more uniform manner,
as indicated by the resource gray scale panel.
This is a directed flocking phase in which
all of the particles move in the same direction,
as shown in the
time series plots of $V_a^x$, $V_a^y$, 
$V_p^x$, and $V_p^y$ in Fig.~\ref{fig:12}(a,b).
Here the system settles into
a state with all particles moving in the positive
$x$ direction.
Additionally, the velocities of the active and passive particles are
almost the same, and
the system moves as a rigid body.
Figure~\ref{fig:10}(d) shows an image of phase V at $N_p/N=0$ and
$\phi=0.88$, where there are only
active particles and a mostly
triangular lattice forms.
In this case, there is some initial
directed flocking motion,
but the system settles into a fluctuating state
that does not exhibit directed motion.
In general, it is necessary to introduce a finite
fraction
of passive particles in order to create 
sufficiently large resource
gradients to induce
directed motion flocking states.
Resources can build up underneath the
passive particles, and can then be consumed by active particles that
displace the passive particles.
Once the flocking motion initiates, it becomes
self sustained since the moving passive particles provide fresh patches of
undepleted resources that attract the active particles through
the resource gradient.
In turn, the active particles push the passive particles
onto new sites which become the next set of resource rich locations.
When the density is high enough,
this directed motion can become locked to
a single direction. At lower densities,
local density fluctuations can break up the flock,
which then reorganizes and moves in a new direction for
a period of time before breaking up again.
When only active particles are present with no passive particles,
the substrate depletion at this resource recovery rate
is much more uniform and flocking motion
does not occur.

In the balanced regime,
within
the intermittent flocking state
at higher densities and lower passive particle
fractions,
we often find that the passive particles have a
higher velocity
than
the active particles.
This is because the clusters into which the active particles
self-assemble often have
zero net velocity.
As an example, consider a system containing 100 particles, of which 70 are
active and 30 are passive. If 40 of the active particles form a cluster with
zero net velocity while the remaining 30 active particles are pushing around
the passive particles, and if the pushed and pushing particles all have the
same velocity $v$, then the average velocity of the active particles is
$V_a=(40 \cdot 0 + 30 \cdot v)/70 \approx 0.43v$, while the average velocity
of the passive particles is $V_p=(30 \cdot v)/30=v$.
Thus on average $V_p>V_a$.
When the passive particle fraction is increased,
the situation is
reversed and the active particles have a higher average velocity
than the passive particles.
In our system there is generally some mixing,
so that even in the regime of large $N_p/N$,
some of the active particles are moving.
As a result, we find that $V_p$ never becomes more than 10\% higher
than $V_a$.
In the scarce regime considered earlier,
the active particles are not mobile enough
to produce self-induced clusters,
so we always find $V_a \geq V_p$.

\section{Plentiful Regime}

\begin{figure}
\includegraphics[width=\columnwidth]{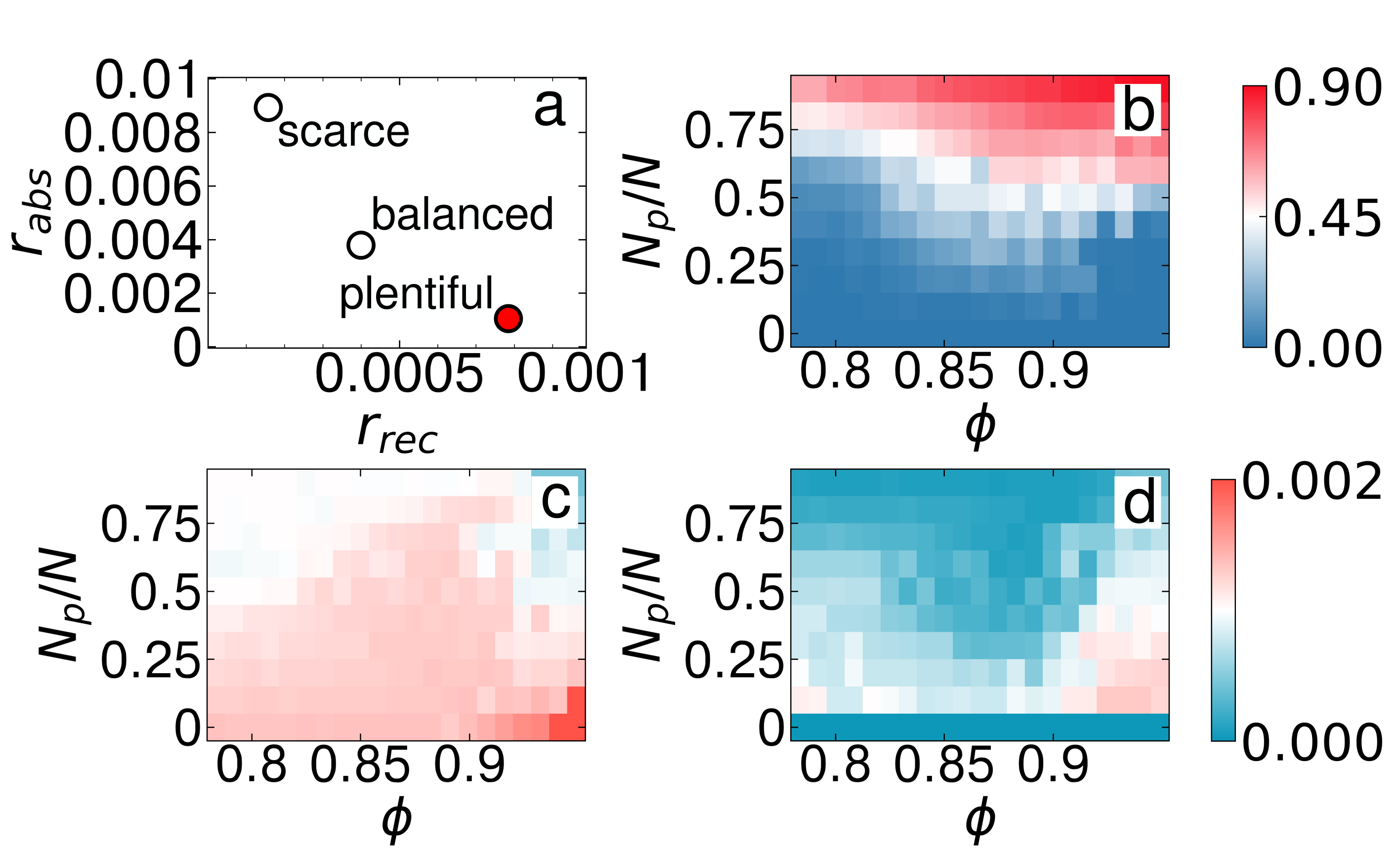}
\caption{(a) Diagram of the location of the scarce ($r_{\rm abs}=0.0087$,
$r_{\rm rec}=0.000175$), balanced ($r_{\rm abs}=0.0036$, $r_{\rm rec}=0.000425$),
and plentiful ($r_{\rm abs}=0.0009$, $r_{\rm rec}=0.000825$) regimes.
(b-d) Results as a function of the fraction of passive particles
$N_p/N$ versus total density $\phi$ from a system in the plentiful
regime, marked by a red dot in panel (a), where the ratio of absorption
to recovery rates is 1.09.
(b) Heat map of $C_L$, the fraction of particles in the largest cluster.
(c) Heat map of $\langle |V_a|\rangle$, the average absolute value of the
active particle velocities.
(d) Heat map of $\langle |V_p|\rangle$, the average absolute value of the
passive particle velocities.}
\label{fig:13}
\end{figure}

In Fig.~\ref{fig:13}(b,c,d) we plot heat maps of $C_L$,
$\langle |V_a|\rangle$, and $\langle |V_p|\rangle$ as a function
of $N_p/N$ versus $\phi$ in the plentiful regime
with $r_{\rm abs}=0.0009$ and $r_{\rm rec}=0.000825$,
indicated by the red dot in Fig.~\ref{fig:13}(a).
When the recovery rate is nearly equal to the absorption rate,
we find larger windows in which flocking in a fixed direction occurs.
Figure~\ref{fig:13}(c) indicates
that the average active particle velocity remains finite even
when $N_p/N = 0$.
In particular, the system with no passive particles now shows
flocking behavior, so we name this state phase V$_a$.
Phase IV still represents directed motion in which
the system acts like a rigid solid moving in a single direction.

\begin{figure}
\includegraphics[width=\columnwidth]{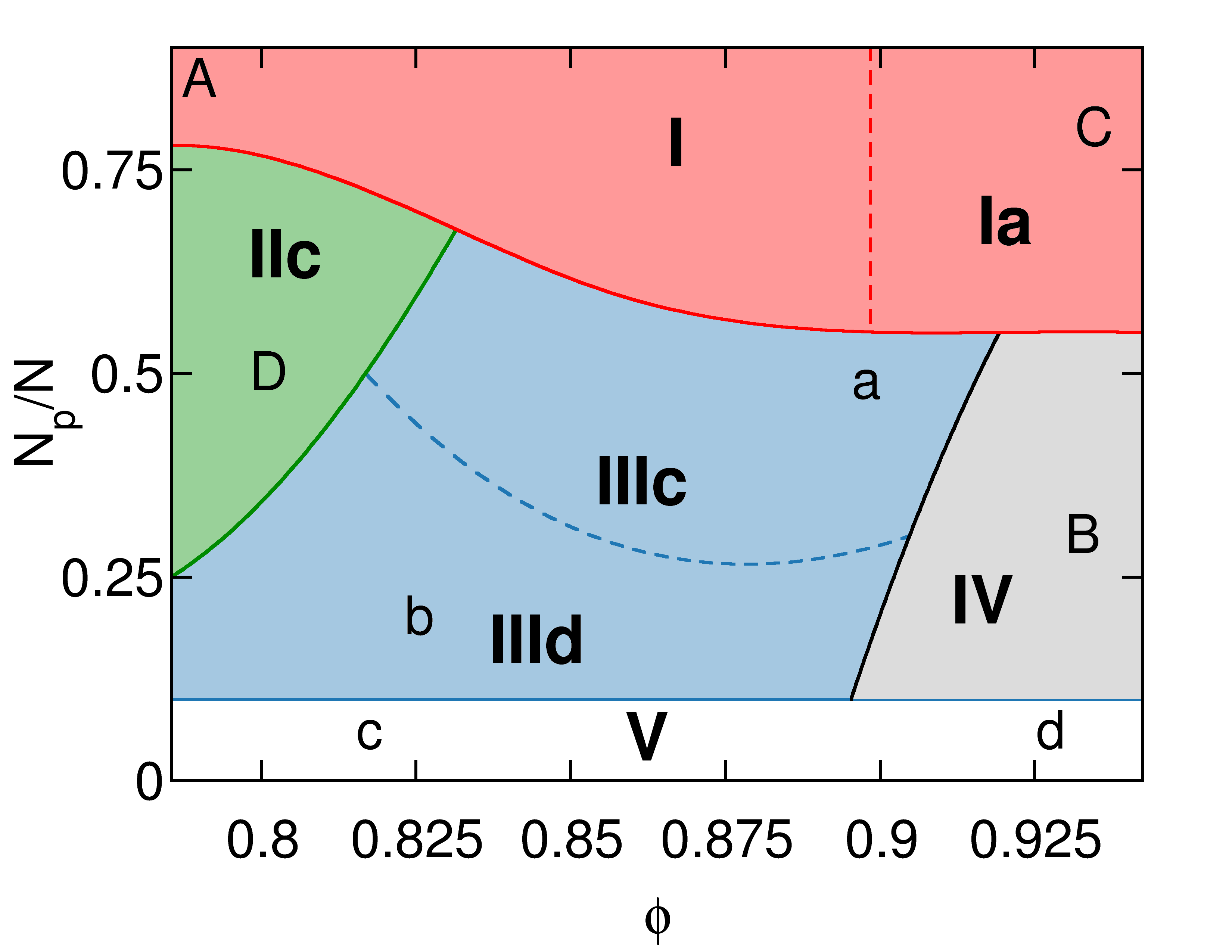}
\caption{Dynamic phase diagram as a function of $N_p/N$ versus $\phi$
for the plentiful regime with $r_{\rm abs} = 0.0009$
and $r_{\rm rec} = 0.000825$.
The phases are: I (phase separated cluster, red left),
I$_{a}$ (mixed jammed state, red right),
III$_{c}$ (flocking active cluster and passive solid, blue top),
III$_d$ (flocking active cluster and fluid passive mixture, blue bottom),
II$_c$ (clustering with no directed motion, green),
IV (directed motion, gray), 
and V$_a$ (no passive particles and directed motion, white).}
\label{fig:14}
\end{figure}

In Fig.~\ref{fig:14} we show the
dynamic phase diagram for the plentiful regime as a function
of $N_p/N$ versus $\phi$.
Phase I is the same
phase separated cluster regime already described in which
the active particles cause the passive particles
to form dense clusters,
while phase I$_a$ is the
same higher density mixed jammed state also found in the balanced regime.
Phase III$_c$ is similar to phase III$_b$ in that the active
particles form clusters,
but now the active clusters
show a flocking behavior
with motion that is 
locked to a single direction,
while the passive particles
form a phase separated solid that has a lower velocity.
In phase III$_d$, flocking active clusters still appear but the
rest of the system is filled with 
a liquid mixture of passive and active particles.
Phase II$_c$ contains clusters of particles that
undergo no directed motion.
In Phase IV, we find the same mixed solid phase with directed
motion as in the scarce and balanced regimes,
and phase V$_a$
contains only active particles
that are undergoing flocking motion.

\begin{figure}
\includegraphics[width=\columnwidth]{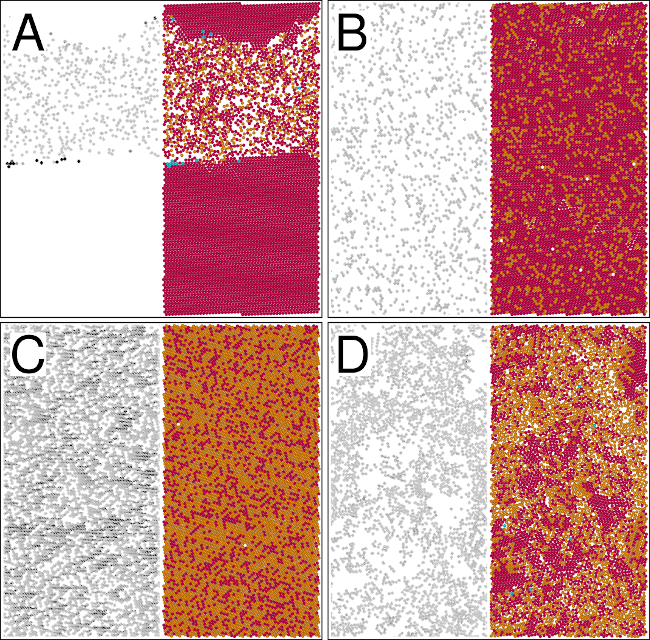}
\caption{Simulation images from the plentiful regime with
$r_{\rm abs}=0.0009$ and $r_{\rm rec}=0.000825$. The left half of each
panel shows a gray scale map of the amount of resource present in the
grid sites, with white indicating $S_g^i=1$ or maximized and black
indicating $S_g^i=0$ or fully depleted. The right half of each panel
shows the positions of the passive particles (red), stationary or
very slowly moving active particles (orange), and active particles that
are moving faster than a threshold velocity (blue). Images correspond to
the points marked A, B, C, and D in the phase diagram of Fig.~\ref{fig:14}.
(A) Image from point A in phase I at
$N_p/N=0.9$ and $\phi =0.785$,
showing a strongly phase separated state.
(B) Image from point B in phase IV
at $N_p/N=0.3$ and
$\phi = 0.935$,
where the system is in a flocking state.
(C) Image from point C in phase I$_a$
at $N_p/N=0.8$ and $\phi = 0.935$.
(d) Image from point D
in phase II$_c$
at $N_p/N=0.8$ and $\phi = 0.801$,
where the system forms a liquid state.
}
\label{fig:15}
\end{figure}

An illustration of
phase I at $N_p/N=0.9$ and $\phi=0.785$ appears in 
Fig.~\ref{fig:15}(A).
The active particles
are confined to lower density areas of the sample
and form a fluctuating
fluid containing a mixture of some passive particles.
Even though the number of active particles is small,
the resource is so plentiful
that the active particles are very mobile and are able to
push the passive particles into a single cluster.
The shepherding of passive particles by active particles
was previously studied
in a heterogeneous system, where it was
shown that when the activity of the active particles is sufficiently
high,
there can be an almost complete phase separation of the active
and passive particles \cite{Forgacs21}.
Other theoretical \cite{Ni14} and
experimental studies \cite{Kummel15,Ramananarivo19} have shown how active matter
particles can assist the crystallization of passive particles.
In these studies,
the rapidly moving active particles 
did not become
trapped inside the passive clusters,
unlike what we observe in the scarce and balanced regimes.
In phase I$_a$, shown in Fig.~\ref{fig:15}(C)
at $N_p/N=0.8$ and $\phi = 0.935$,
the active and passive particles are mixed
and form a uniform jammed triangular solid where there is little motion.
Figure~\ref{fig:15}(B) illustrates phase IV
at $N_p/N=0.3$ and $\phi = 0.935$.
A uniform solid appears and moves in a flocking
fashion along the $x$ direction.
Phase II$_c$ at $N_p/N=0.8$ and $\phi=0.801$,
shown in Fig.~\ref{fig:15}(D),
has some localized clustering of passive particles,
but the active particles are moving rapidly 
enough to break up the passive clusters and there is no
coordinated directed
motion. There is, however, directed motion of the clusters over
short times, with each cluster moving in a particular direction for
a brief period of time before rearranging and moving in a new
direction for another brief period of time.

\begin{figure}
\includegraphics[width=\columnwidth]{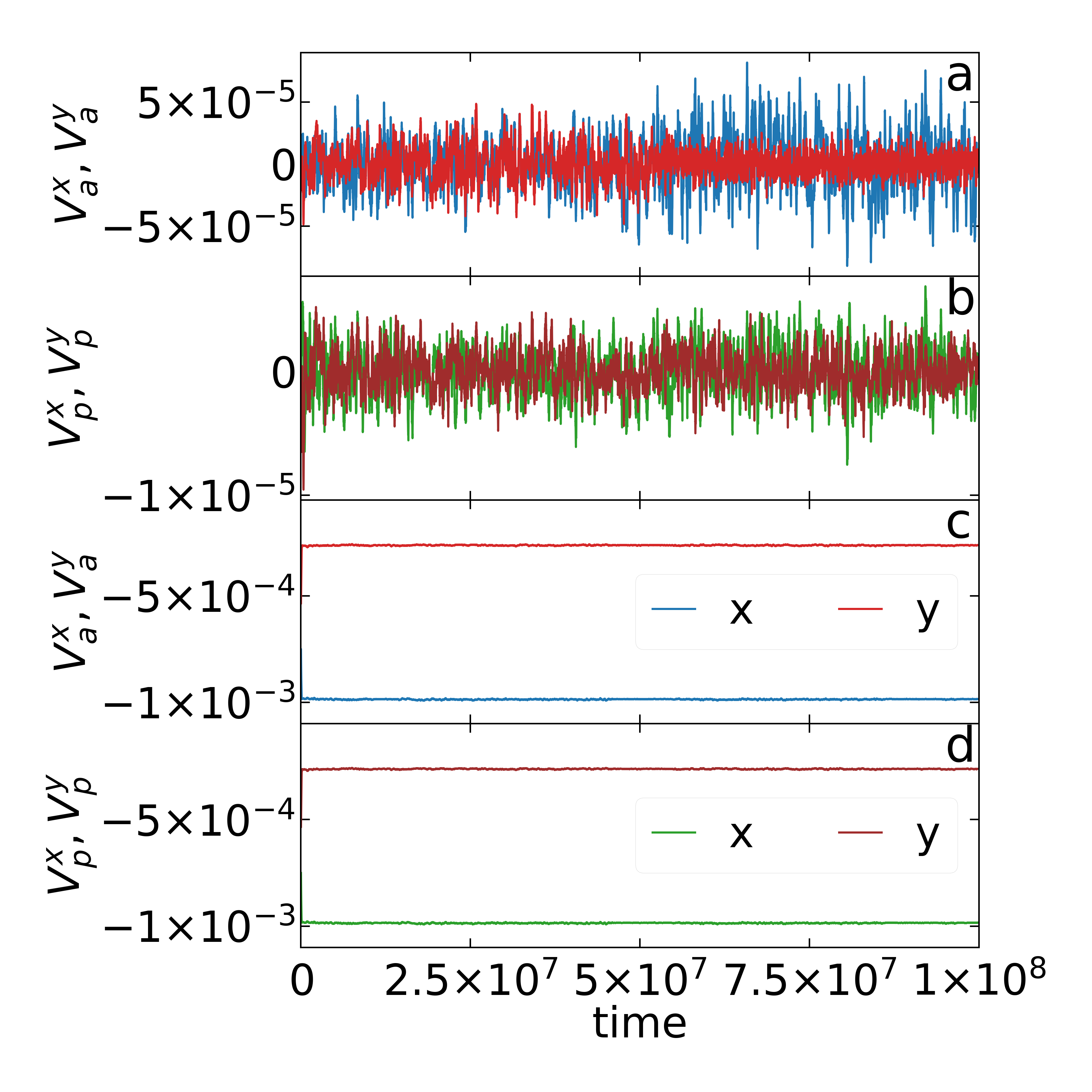}
\caption{
Time series of
the active particle velocities $V_a^x$ (blue)
and $V_a^y$ (red)
and the passive particle velocities $V_p^x$ (green)
and $V_p^y$ (maroon)
in the plentiful regime with $r_{\rm abs}=0.0009$ and
$r_{\rm rec}=0.000825$ at labeled points from the phase diagram
of Fig.~\ref{fig:14}.
(a) $V_a^x$ and $V_a^y$ in phase I at point A with
$N_p/N=0.9$ and $\phi=0.785$, illustrated in
Fig.~\ref{fig:15}(A).
(b) The corresponding
$V_p^x$
and $V_p^y$.
In this case there is no directed motion.
(c)
$V_a^x$
and $V_a^y$
in phase
IV
from point B with
$N_p/N=0.3$ and $\phi=0.935$, illustrated
in Fig.~\ref{fig:15}(B).
(d) The corresponding
$V_p^x$
and $V_p^y$.
Here there is directed motion with both species moving in the same
direction.
}
\label{fig:16}
\end{figure}

In Fig.~\ref{fig:16}(a,b) we
plot time series of the velocities $V_a^x$, $V_a^y$ and $V_p^x$, $V_p^y$ for the
active and passive particles, respectively,
for phase I from Fig.~\ref{fig:15}(A).
The velocities are fluctuating around zero,
and although the active particles
show much stronger velocity fluctuations than the passive particles,
no flocking motion occurs.
In phase IV from Fig.~\ref{fig:15}(B),
the time series plots of $V_a^x$, $V_a^y$, $V_p^x$, and $V_p^y$
in Fig.~\ref{fig:16}(c,d)
indicate that both species
are moving as a solid at the same velocity.
In this case,
the particles are moving in the negative $x$ and negative $y$
directions, giving 
flow along an angle of $-105^\circ$ from the $x$ direction.
In general, the flocks can organize to
move in any direction, similar to the Vicsek 
flocking models  \cite{Vicsek12,Morin17}.

\begin{figure}
\includegraphics[width=\columnwidth]{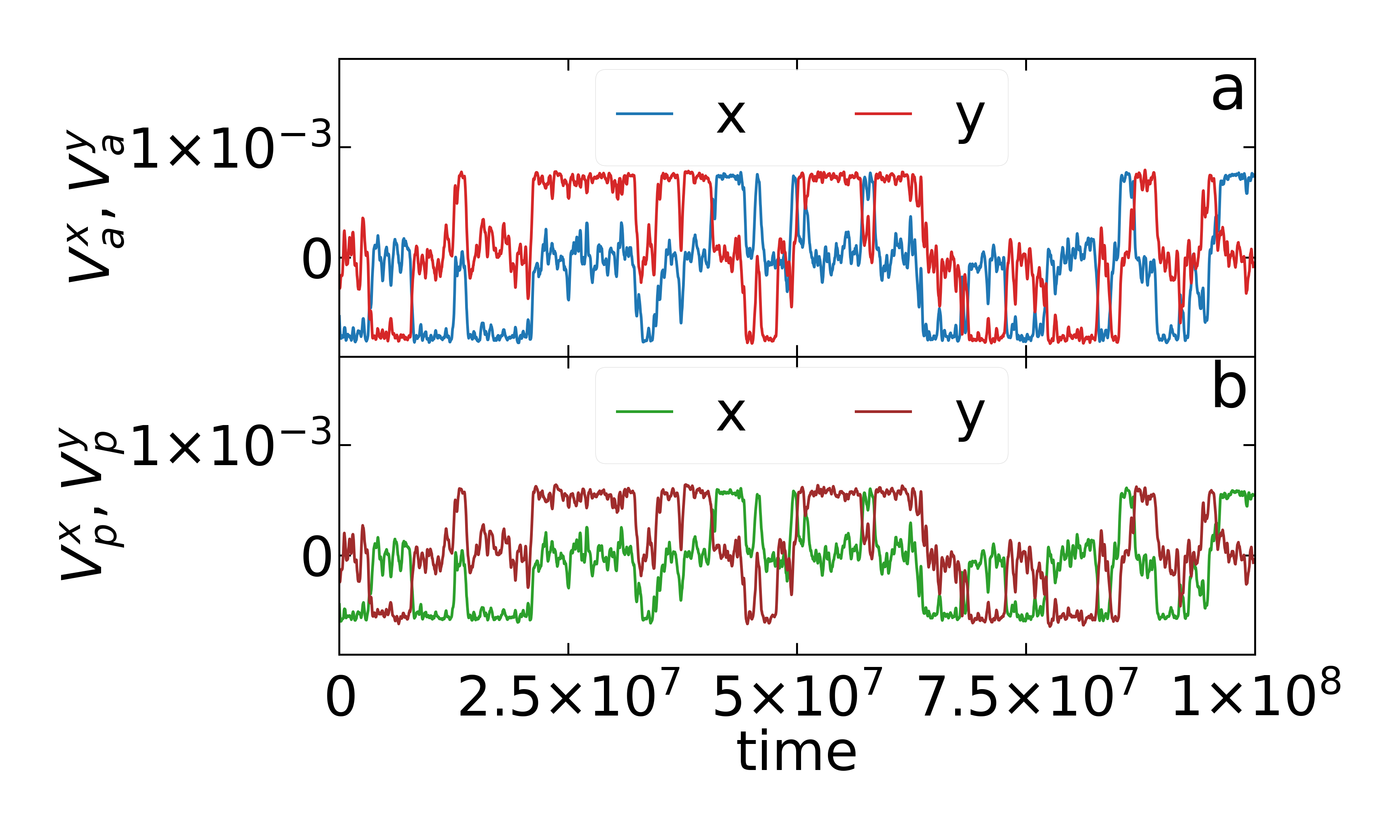}
\caption{
(a) Time series $V_a^x$ (blue) and $V_a^y$ (red) for the active particles
in the plentiful regime with $r_{\rm abs}=0.0009$ and $r_{\rm rec}=0.000825$
for phase II$_c$ at point D in
the phase diagram of Fig.~\ref{fig:14} with
$N_p/N=0.8$ and $\phi=0.801$,
illustrated in Fig.~\ref{fig:15}(D),
showing switching directed motion.
(b) The corresponding time series $V_p^x$ (blue) and $V_p^y$ (maroon) for
the passive particles
showing that they
are
being fully entrained by the active particles.
}\label{fig:17}
\end{figure}

Figure~\ref{fig:17}(a,b) shows time series of $V_a^x$, $V_a^y$,
$V_p^x$, and $V_p^y$ 
in phase II$_c$ from Fig.~\ref{fig:15}(D). Here there
is intermittent flocking
of mixtures of both passive and active particles, where the
direction of motion of the flocks makes abrupt changes from time
to time.

\begin{figure}
\includegraphics[width=\columnwidth]{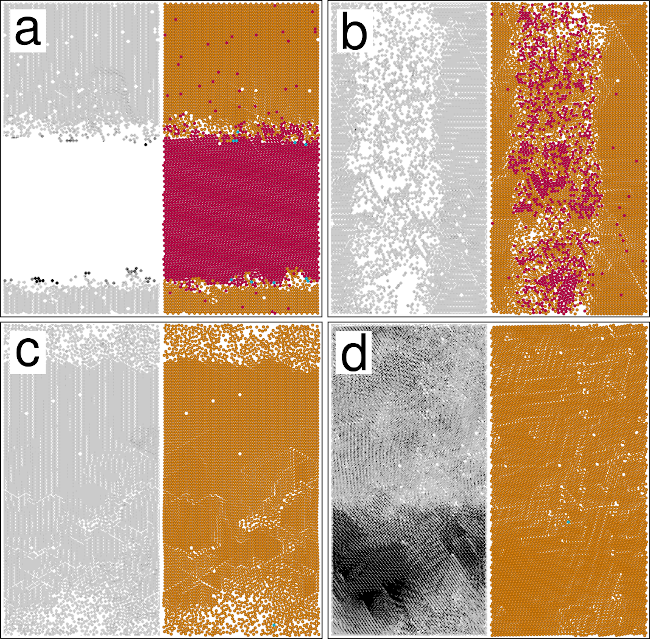}
\caption{
Simulation images from the plentiful regime with $r_{\rm abs}=0.0009$ and
$r_{\rm rec}=0.000825$. The left half of each panel shows a gray scale map
of the amount of resource present in the grid sites, with white indicating
$S_g^i=1$ or maximized and black indicating $S_g^i=0$ or fully depleted.
The right half of each panel shows the positions of the passive particles
(red), stationary or very slowly moving active particles (orange), and
active particles that are moving faster than a threshold velocity (blue).
Images correspond to the points marked a, b, c, and d in the phase diagram
of Fig.~\ref{fig:15}.
(a) Image from point a in phase III$_c$
at $N_p/N=0.5$ and $\phi = 0.895$.
(b) Image from point b in phase III$_d$
at $N_p/N=0.2$ and
$\phi = 0.825$.
(c) Image from point c
in phase V$_a$ at $N_p/N=0$
and $\phi = 0.812$.
(d) Image from point d
in phase V$_a$ at $N_p/N=0$ and $\phi = 0.927$.
}
\label{fig:18}
\end{figure}

In phase III$_c$,
there is a phase separation into
passive and active crystalline states where the active particles exhibit
directed motion and the passive particles
form an ordered crystal state at
higher densities, as shown in Fig.~\ref{fig:18}(a)
for $N_p/N=0.5$ and $\phi = 0.895$.
Phase III$_d$ is an active crystal state containing 
liquid like passive particles,
illustrated in Fig.~\ref{fig:18}(b) for $N_p/N=0.2$ and
$\phi = 0.825$.
The velocity difference
between the two species is higher in phase III$_c$ than
in phase III$_d$
since the passive solid can
decouple from the active solid
more effectively than the passive liquid can.

\begin{figure}
\includegraphics[width=\columnwidth]{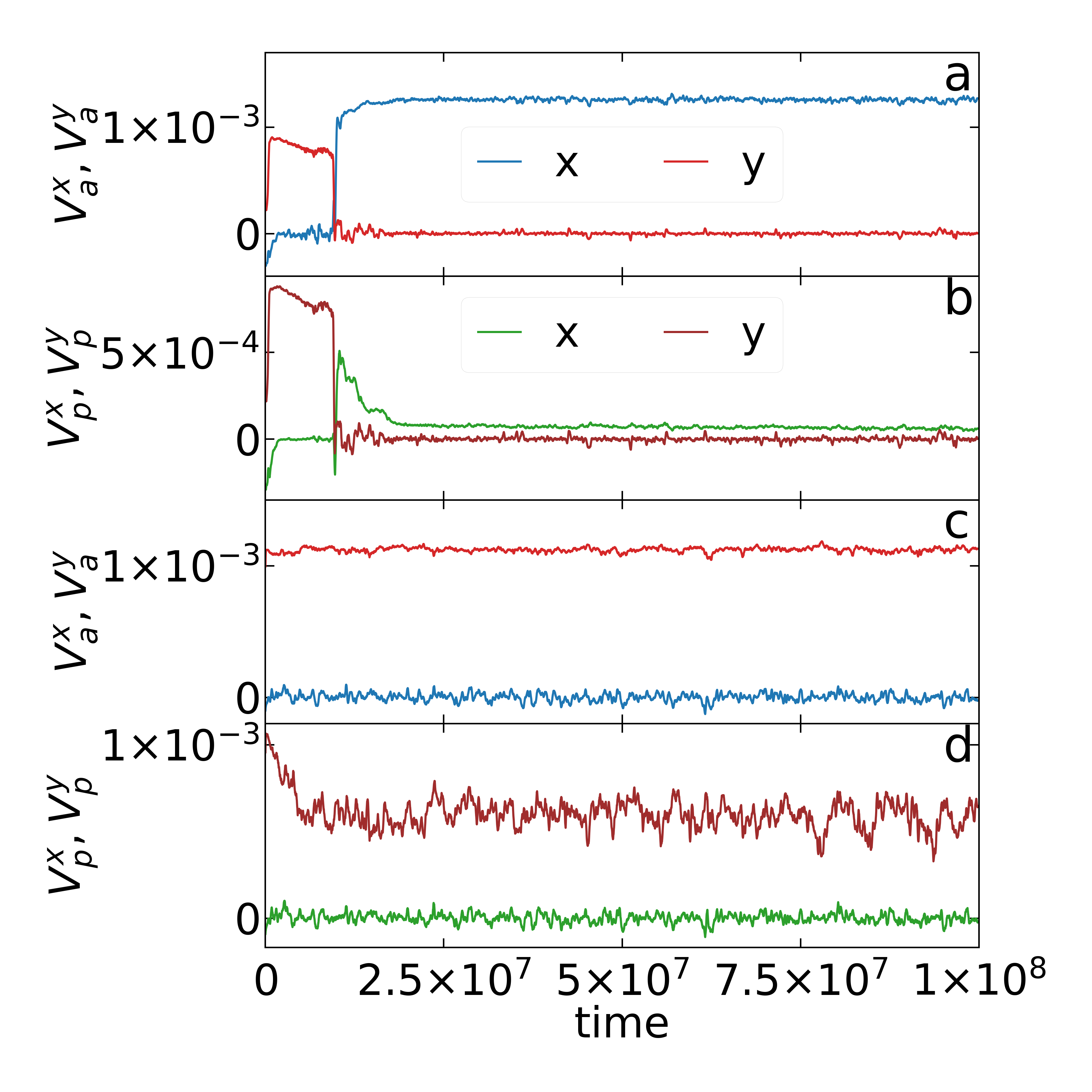}
\caption{
Time series $V_a^x$ (blue) and $V_a^y$ (red) for the active
particles
and $V_p^x$ (green) and $V_p^y$ (maroon) for the passive
particles
in the plentiful regime with $r_{\rm abs}=0.0009$ and
$r_{\rm rec}=0.000825$ at labeled points from the phase diagram
of Fig.~\ref{fig:14}.
(a) $V_a^x$ and $V_a^y$ in phase III$_c$ at point a
with $N_p/N=0.5$ and $\phi=0.895$,
illustrated in Fig.~\ref{fig:18}(a).
(b) The corresponding
$V_p^x$
and $V_p^y$.
In this phase, the active particles are flocking but
the passive particles show almost no directed motion.
(c)
$V_a^x$
and $V_a^y$
in phase III$_d$ from point b
at $N_p/N=0.2$ and $\phi=0.825$, illustrated in 
Fig.~\ref{fig:18}(b).
(d) The corresponding
$V_p^x$
and $V_p^y$
for the passive particles.
In this phase,
the active particles are flocking and
the passive particles show limited directed motion.
}\label{fig:19}
\end{figure}

In Fig.~\ref{fig:19}(a,b) we plot
time series of the active and passive velocities
$V_a^x$, $V_a^y$, $V_p^x$, and $V_p^y$
for phase III$_c$ from Fig.~\ref{fig:18}(a).
Over time the system
organizes to a state where the active particles are moving
with a large velocity in the $x$ direction
while the passive particles are barely moving and have
only a small velocity along 
the $x$ direction due to the entrainment of 
some of the passive particles
by the active particles.
Fig.~\ref{fig:19}(c,d) shows similar velocity time series for
phase III$_b$ from
Fig.~\ref{fig:18}(b).
Here $V_a^y$ is finite and $V_a^x$ is close to zero
since the active particles have organized into
motion along the positive $y$-direction.
The passive particles are also moving on average in
the positive $y$ direction but have a lower
velocity than the active particles, since
some of the passive particles are intermixed with the active cluster and
are dragged along by it.

\begin{figure}
\includegraphics[width=\columnwidth]{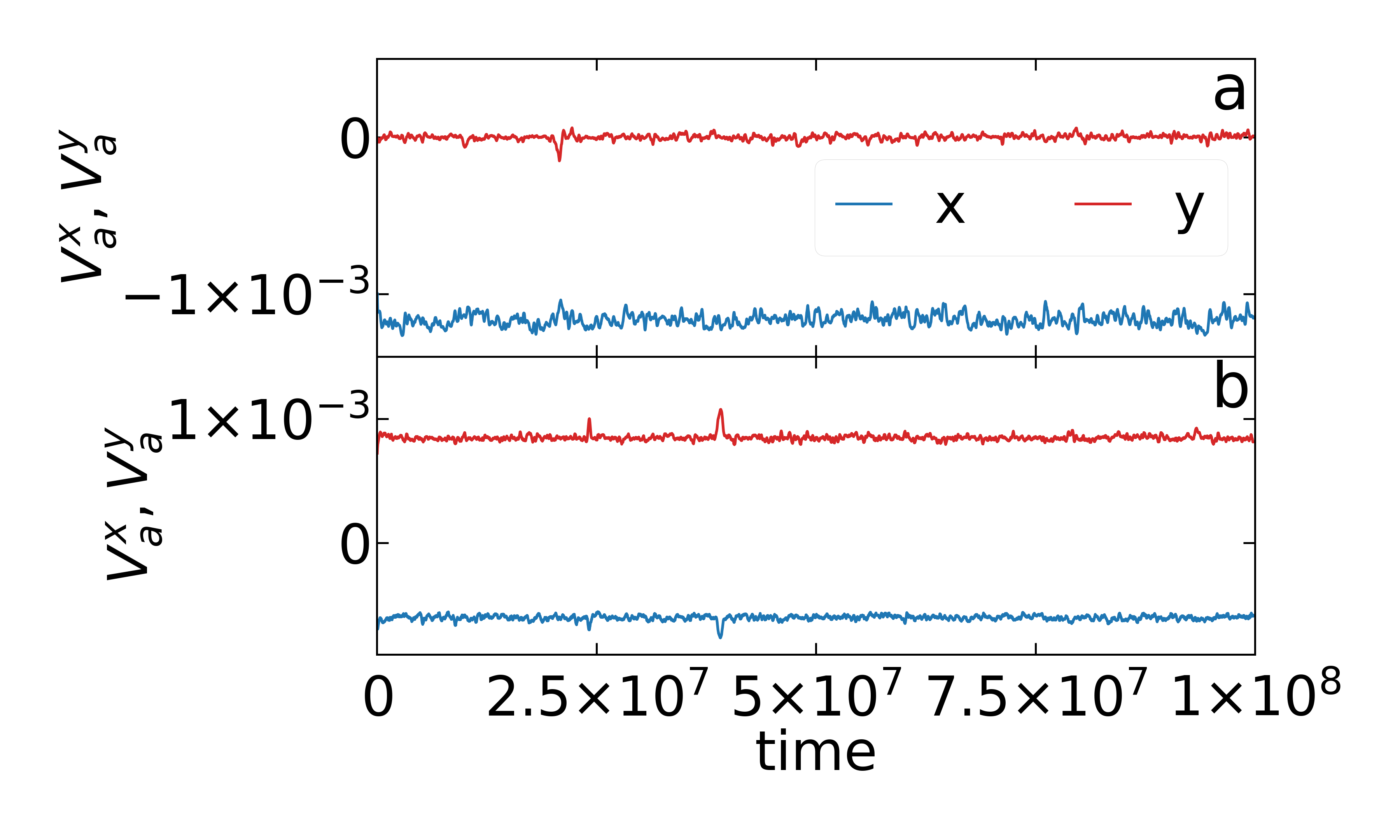}
\caption{
Time series $V_a^x$ (blue) and $V_a^y$ (red) for the active particles
in the plentiful regime with $r_{\rm abs}=0.0009$ and $r_{\rm rec}=0.000825$
in phase $V_a$ from the labeled points in the phase diagram
of Fig.~\ref{fig:14} for
$N_p/N=0$, where passive particles are absent.  
(a)
Point c with $\phi=0.812$, illustrated in Fig.~\ref{fig:18}(c),
showing flocking.
(b) Point d with $\phi=0.927$, illustrated in Fig.~\ref{fig:18}(d),
also showing flocking.
}\label{fig:20}
\end{figure}

An illustration of phase V$_a$
at $N_p/N=0$ and
$\phi = 0.812$ appears in
Fig.~\ref{fig:18}(c).
There is a large active cluster with triangular ordering
surrounded by a lower density fluid,
and the system exhibits directed  motion
as indicated by the time series plots of $V_a^x$ and $V_a^y$ in
Fig.~\ref{fig:20}(a).
Figure~\ref{fig:18}(d) shows phase V$_a$
for $N/N_p=0$ at a higher density of
$\phi = 0.927$,
where a mostly crystalline jammed solid moves at an angle to the $y$
direction.
The gradient in the resource substrate is clearly visible in the
gray scale panel of Fig.~\ref{fig:18}(d).
The active particle velocities $V_a^x$ and $V_a^y$ for
the system in Fig.~\ref{fig:18}(d) with $\phi=0.927$ appear
in Fig.~\ref{fig:20}(b),
where it is clear that the particles are moving at an angle to both the
$x$ and $y$ directions.
The flocking phases generally appear only above
a critical particle density and a critical resource recovery rate.
At high densities,
the system generally can reach
flocking configuration rather rapidly,
but as the density is reduced, the period of fluctuating
motion that precedes the emergence of a flocking state
becomes longer and longer.

\begin{figure}
\includegraphics[width=\columnwidth]{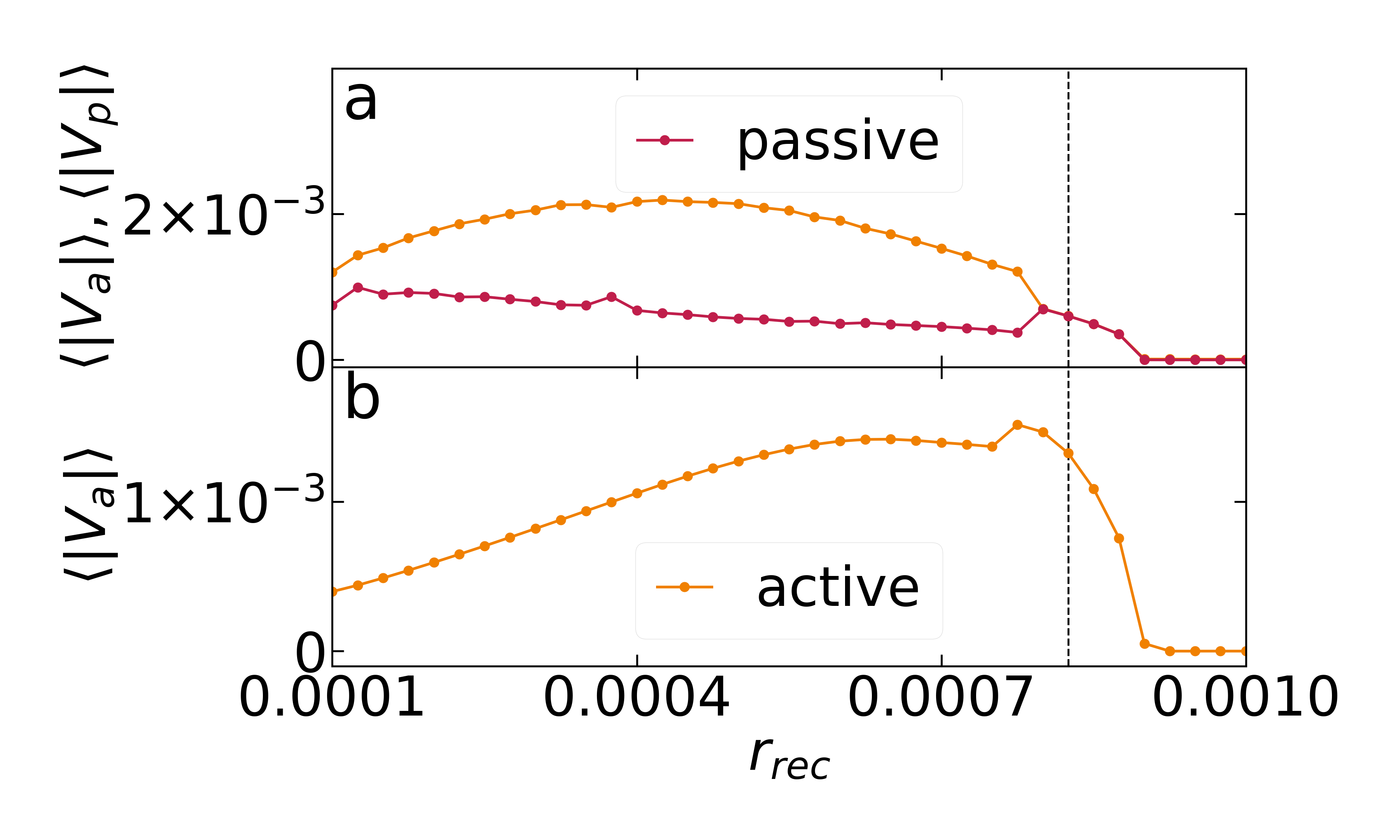}
\caption{(a) Average absolute value $\langle |V_a|\rangle$ (orange) of the
active particles and $\langle |V_p|\rangle$ (red) of the passive particles vs
$r_{\rm rec}$
for
a sample with $\phi = 0.935$, $r_{\rm abs} = 0.0009$,
and $N_p/N = 0.8$.
The dashed line corresponds to the
value of $r_{\rm rec}$ at point C of the phase diagram in Fig.~\ref{fig:14},
where the system is phase I$_a$.
There is a transition from phase I$_a$ to
phase III$_c$ at $r_{\rm rec} = 0.0008$.
(b) $\langle |V_a|\rangle$ vs $r_{\rm rec}$
for a sample with $r_{\rm abs} = 0.0009$, $\phi = 0.817$ and $N_{p}/N = 0.0$,
where only active particles are present and flocking behavior appears.
The dashed line corresponds to point c from the phase diagram in
Fig.~\ref{fig:14}, where the system is in phase V$_a$.
The cusp just below $r_{\rm rec} = 0.0008$
falls at the transition from
flocking in changing directions
for lower $r_{\rm rec}$ to flocking in
a fixed direction for higher $r_{\rm rec}$.
}\label{fig:21}
\end{figure}

We can also examine the evolution of the flocking phase
by fixing the density
and varying the recovery rate.
In Fig.~\ref{fig:21}(a)
we plot the average
absolute value $\langle |V_a|\rangle$ and $\langle |V_p|\rangle$
of the active and passive particles
versus $r_{\rm rec}$ for a system with $\phi = 0.935$,
$r_{\rm abs} = 0.0009$, and $N_p/N = 0.8$.
The dashed line corresponds to point C in the phase diagram of
Fig.~\ref{fig:14},
where the system is
in the jammed low velocity mixed solid phase I$_a$,
which is stable over the range
$0.0008 < r_{\rm rec} < 0.0009$.
For $r_{\rm rec} > 0.0009$,
the velocity of both passive and active particles drops nearly to
zero since
the resources recover rapidly enough that the active
particles can continuously consume resources from the grid sites
without building up a resource gradient.
For $0.00015 < r_{\rm rec} < 0.0008$,
the active particles form a flocking solid that
is phase separated from the passive particles,
so the behavior is similar to that found in
phase III$_c$.
When $r_{\rm rec} < 0.00015$,
resource depletion becomes an issue and the motion becomes
pulse-like. Here the
flocking is lost and the behavior resembles that of
phase II$_c$.
Overall, the flocking produces a nonmonotonic $\langle |V_a|\rangle$,
indicating that there is an optical
recovery rate at which the largest flocking velocities occur.
This is due to a competition between excessive resource depletion
for smaller $r_{\rm rec}$
and disappearance of the resource gradient for larger $r_{\rm rec}$.
Across the III$_c$ to I$_a$ transition, the velocity
of the active particles drops
but that of the passive particles increases as the
motion of the two species
becomes locked together.
It is beyond the scope of this work to determine whether the change
from phase III$_c$ to
phase I$_a$ is a true phase transition;
however, this would be an interesting future direction to
explore.

In Fig.~\ref{fig:21}(b)
we plot $\langle |V_a|\rangle$
versus $r_{\rm rec}$
for a system with $r_{\rm abs} = 0.0009$, $\phi = 0.817$,
and $N_{p}/N = 0.0$, where only active particles are present.
The dashed line corresponds to point c
in the phase diagram of Fig.~\ref{fig:14} where, as shown
in Fig.~\ref{fig:18}(c),
the system forms a crystalline solid
coexisting with a lower density fluid.
A flocking solid appears for
$0.00075 < r_{\rm rec} < 0.0009$.
For $r_{\rm rec} > 0.0009$, motion ceases when the resource recovery
becomes so rapid that resource gradients never form.
For $ r_{\rm rec} < 0.00075$,
the flocking is no longer directed but becomes intermittent, and the
flocking direction changes repeatedly with relatively brief time periods
spent flocking in any one direction.
As the recovery rate decreases further,
the switching between flocking directions becomes more frequent
and the system eventually enters the pulse like flow regime
studied previously \cite{Varga22}.
Earlier work on flocking models also showed that
there can be an optimal amount of
disorder
that produces the greatest amount of collective motion \cite{Chepizhko13}.

\begin{figure}
\includegraphics[width=\columnwidth]{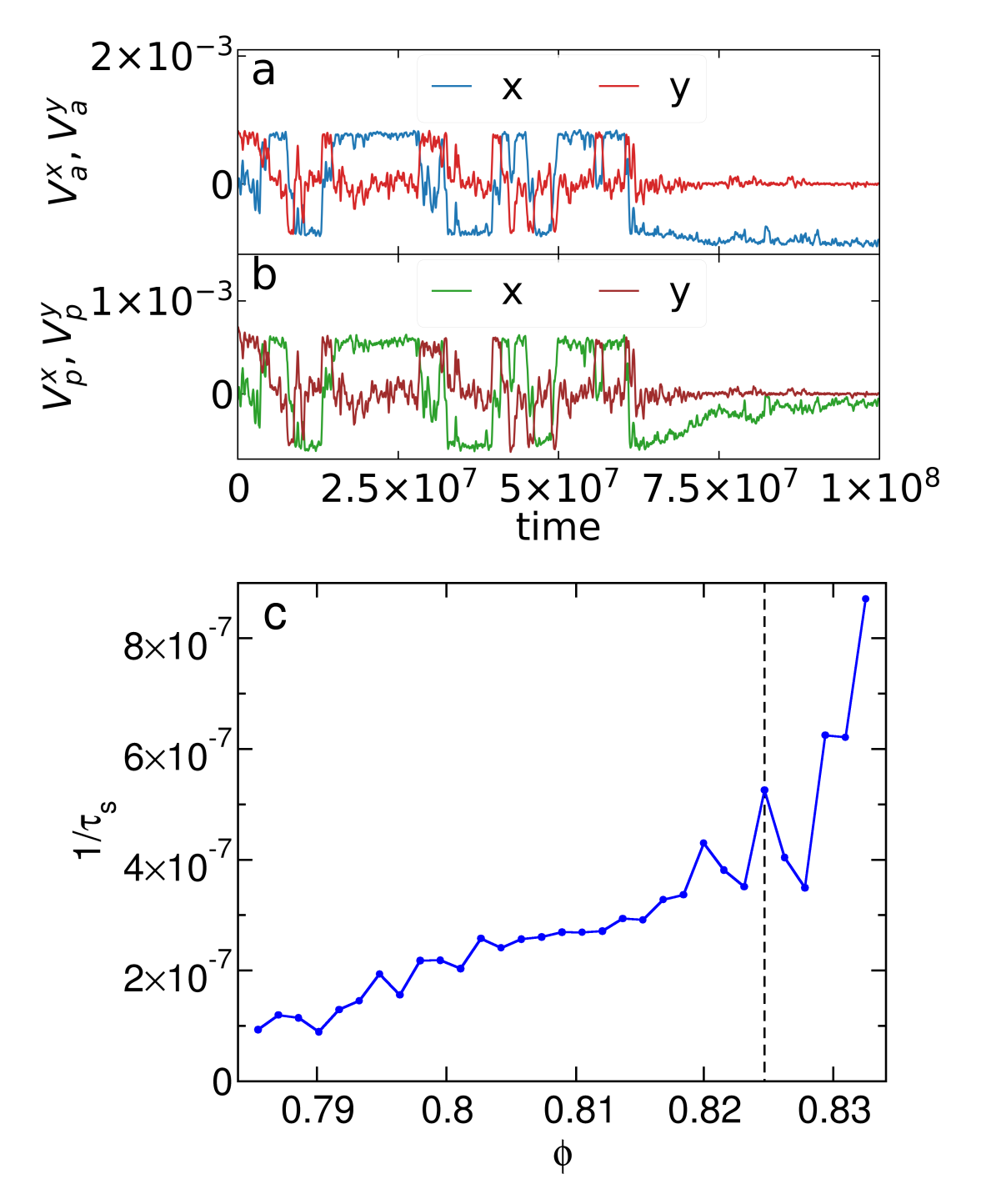}
\caption{
(a) Time series $V_a^x$ (blue) and $V_a^y$ (red) for the active particles in
the plentiful regime with $r_{\rm abs}=0.0009$ and $r_{\rm rec}=0.000825$ 
for $N_p/N=0.5$ and $\phi=0.82$,
at the II$_c$ to III$_c$ transition. The system
settles into a directed motion state at long times.
(b) The corresponding time series of $V_p^x$ (green) and $V_p^y$ (maroon) for
the passive particles.
(c) $1/\tau_s$, the
inverse of the average time interval between flocking direction
switches in the transient regime of the system in panels (a) and
(b), vs $\phi$.
}\label{fig:22}
\end{figure}

We also more closely explored the transition
from flocking in
repeatedly changing directions to flocking in a single direction.
In Fig.~\ref{fig:22}(a,b) we plot time series of $V_a^x$, $V_a^y$,
$V_p^x$, and $V_p^y$ 
for a system in the plentiful regime at
$N_{p}/N = 0.5$ and $\phi = 0.82$,
on the boundary between phase II$_c$ and phase III$_c$.
During the first $7 \times 10^7$ simulation time steps,
the system is in a flocking state that
moves in some direction for
a period of time before switching to a new direction for another
period of time, and the active and passive particles are mixed
together.
After this long transient interval, the system settles into
phase III$_c$ where the active and passive particles decouple.
Here
the active particles are flocking along the $-x$ direction
while the passive particles form a cluster
or liquid state with no net translation.
In general, as $\phi$ approaches the II$_c$-III$_c$
boundary,
it takes longer for the system
to settle into a directed motion state,
and at some critical density the system
remains in the fluctuating phase II$_c$ state indefinitely.

We measure the average time interval $\tau_s$ that the system
in Fig.~\ref{fig:22}(a,b) spends moving in a given direction
before switching to a new direction during the transient interval
preceding the onset of
III$_c$ directed motion.
In Fig.~\ref{fig:22}(c) we plot $1/\tau_s$ as a function of $\phi$.
As $\phi$ increases, $1/\tau_s$ becomes
larger,
suggesting that there is a critical density
$\phi_{c}$ at which the switching time vanishes
and the system forms a coherent flock.
The data is not accurate enough to determine
whether there is a power law
divergence of $1/\tau_s$,
but the results suggest that there is a critical density
above which
a directed motion flocking state appears.  More accurate
data or larger systems would be required to learn whether
this transition
falls into the continuous Vicsek class
or
whether this is an example of
an absorbing phase transition \cite{Hinrichsen00,Corte08,Reichhardt14a},
where the switching flocking behavior is the fluctuating state
and the directed flocking motion is the absorbing state.

\section{Discussion}
Our work describes a distinctive class of active matter that
differs from most self-driven particle based systems,
where each particle is subjected to a fixed driving force.
In our system, the
forces that generate the motion can vary strongly
both from particle to particle and over time
because they depend on the competing resource
absorption and recovery processes.
Despite this fact, we still observe several of the same
features found in
standard particle based active matter systems.
For example, when the forces producing the motion of the active
particles become high enough,
there can be a motility induced phase separation of the active particles,
or the active particles can
induce a phase separation of the passive particles.
In addition, our model captures
features commonly observed in flocking systems, such
as the formation
of coherently moving states that select a symmetry-broken
direction of motion.
This occurs even though
our model does not 
include an alignment term for the motion of neighboring particles,
which is typically required to produce
flocking in Vicsek models.
We also note that the directed
motion we observe is different from that
found in active ratchet
systems \cite{Wan08,Reichhardt17a,Borba20,Obyrne22},
where the active particles undergo directed motion when  coupled to
an asymmetric substrate.
In our case, the substrate is symmetric,
but a dynamically generated asymmetry can arise in the resource
gradient due to local density fluctuations of the active and passive
particles.

Possible realizations of the model we consider include
the robotic assembly on an optical resource landscape
that motivated this work \cite{Wang21}.
Our results could
also be of relevance
to a number of biological systems, such as
crawling cells that exhibit heptotaxis by generating and then
following a gradient 
\cite{Weber13,Tweedy16,Dona13},
modifying the surface on which they are moving \cite{Solon06},
or have their
activity controlled by
their rate of resource consumption \cite{CochetEscartin21}.
There are also several models
in which
active particles interact with their own secreted trails
\cite{Sengupta09} or undergo
synthetic chemotaxis \cite{Liebchen18}.
In the case of artificial swimmers,
for colloids that react with a surface,
the surface could regain fuel or resources over time,
or the colloids could be coupled to
an optical feedback that could be used to generate effective
rules of motion \cite{Lavergne19,Yang21}.
Our results could also be useful for systems near jamming
\cite{Bi16},
such as biological assemblies containing
cells with different mobilities.
Many other variations of this model could be
explored
in the future, including:
adding longer range interactions;
introducing spatial
heterogeneity by
setting certain resource sites to be always full or always
empty or setting the maximum resource level of a portion of the
sample to a lower level;
having more than two species of particles;
introducing additional types of noise;
or
providing rules by which active particles can transform into passive
particles over time and vice versa.

\section{Summary}

We have introduced a model
of a system of particles that move on a resource landscape
grid.
The system contains both active particles, which
consume the resources on which they are sitting at a certain rate,
and passive particles, which do not consume resources.
The active particles experience a force from
the substrate that is proportional to the
local resource gradient, causing them
to move toward sites containing higher resource levels.
All particles, both active and passive, interact sterically with each other.
We consider several different ratios of the resource absorption and
recovery rates
and also vary the
particle density and the ratio of passive to active particles.
Despite the simplicity of this model,
we show that it exhibits a remarkable
number of phases, including motility 
induced phase separated states, clustering of the active particles,
and directed flocking motion.
In the scarce regime where resource absorption is dominant,
we find different types of mixed phases and riverlike 
phase separated flows, while
at higher densities, mixed jammed states appear.  
For the balanced regime where the resource absorption rate is
about four times higher than the recovery rate,
we observe
a wealth of phase separated regimes
with different morphologies,
including stripe, bubble, and void states.
The activity in the balanced regime
is high enough that the active particles
are able to self-cluster and form
a motility induced phase separated state.
As the fraction of passive particles increases,
the mobility of the active particles can increase
due to the formation of
large resource gradients at the edges of the passive
particle clusters.
We also find a directed flocking motion regime
in which the particles all move in a symmetry broken direction
due to the generation of a self-sustaining resource gradient.
This directed motion resembles the
flocking states found in Vicsek models, and it appears in
systems containing
only active particles
as well as for mixtures of 
active and passive particles
at
high densities where a jammed solid forms.
In the plentiful regime where the absorption and recovery rates
are nearly equal,
we find an increased number of flocking regimes.
Here the phase separation can become
so strong
that the active particles decouple from the passive particles. 
The flocking mobility
varies nonmonotonically with
changing resource recovery rate.
For the highest recovery rates,
there is no motion since resource gradients never form, while
at low recovery rates,
there is only a low level of
intermittent motion
since sizable gradients only slowly form.
As a result,
there is an optimal recovery rate for
generating the greatest amount of directed mobility. 
In general, there is a transient period in which the flocking motion
repeatedly switches directions
before
eventually settling down into motion in a single direction.
The time interval between direction switches decreases as the density of
the system increases.
We discuss how our system could be relevant for various types
of biological or robotic systems, colloids on 
reactive substrates, or colloids with optical feedback control. 

\begin{acknowledgments}
This work was supported by the US Department of Energy through
the Los Alamos National Laboratory.  Los Alamos National Laboratory is
operated by Triad National Security, LLC, for the National Nuclear Security
Administration of the U. S. Department of Energy (Contract No. 892333218NCA000001).
LV and AL were supported by a grant of the Romanian Ministry of Education
and Research, CNCS - UEFISCDI, project number
PN-III-P4-ID-PCE-2020-1301, within PNCDI III.
\end{acknowledgments}

\bibliography{mybib}

\begin{thebibliography}{55}%
\makeatletter
\providecommand \@ifxundefined [1]{%
 \@ifx{#1\undefined}
}%
\providecommand \@ifnum [1]{%
 \ifnum #1\expandafter \@firstoftwo
 \else \expandafter \@secondoftwo
 \fi
}%
\providecommand \@ifx [1]{%
 \ifx #1\expandafter \@firstoftwo
 \else \expandafter \@secondoftwo
 \fi
}%
\providecommand \natexlab [1]{#1}%
\providecommand \enquote  [1]{``#1''}%
\providecommand \bibnamefont  [1]{#1}%
\providecommand \bibfnamefont [1]{#1}%
\providecommand \citenamefont [1]{#1}%
\providecommand \href@noop [0]{\@secondoftwo}%
\providecommand \href [0]{\begingroup \@sanitize@url \@href}%
\providecommand \@href[1]{\@@startlink{#1}\@@href}%
\providecommand \@@href[1]{\endgroup#1\@@endlink}%
\providecommand \@sanitize@url [0]{\catcode `\\12\catcode `\$12\catcode
  `\&12\catcode `\#12\catcode `\^12\catcode `\_12\catcode `\%12\relax}%
\providecommand \@@startlink[1]{}%
\providecommand \@@endlink[0]{}%
\providecommand \url  [0]{\begingroup\@sanitize@url \@url }%
\providecommand \@url [1]{\endgroup\@href {#1}{\urlprefix }}%
\providecommand \urlprefix  [0]{URL }%
\providecommand \Eprint [0]{\href }%
\providecommand \doibase [0]{http://dx.doi.org/}%
\providecommand \selectlanguage [0]{\@gobble}%
\providecommand \bibinfo  [0]{\@secondoftwo}%
\providecommand \bibfield  [0]{\@secondoftwo}%
\providecommand \translation [1]{[#1]}%
\providecommand \BibitemOpen [0]{}%
\providecommand \bibitemStop [0]{}%
\providecommand \bibitemNoStop [0]{.\EOS\space}%
\providecommand \EOS [0]{\spacefactor3000\relax}%
\providecommand \BibitemShut  [1]{\csname bibitem#1\endcsname}%
\let\auto@bib@innerbib\@empty
\bibitem [{\citenamefont {Bhattacharya}\ and\ \citenamefont
  {Higgins}(1993)}]{Bhattacharya93}%
  \BibitemOpen
  \bibfield  {author} {\bibinfo {author} {\bibfnamefont {S.}~\bibnamefont
  {Bhattacharya}}\ and\ \bibinfo {author} {\bibfnamefont {M.~J.}\ \bibnamefont
  {Higgins}},\ }\bibfield  {title} {\enquote {\bibinfo {title} {Dynamics of a
  disordered flux line lattice},}\ }\href {\doibase
  10.1103/PhysRevLett.70.2617} {\bibfield  {journal} {\bibinfo  {journal}
  {Phys. Rev. Lett.}\ }\textbf {\bibinfo {volume} {70}},\ \bibinfo {pages}
  {2617--2620} (\bibinfo {year} {1993})}\BibitemShut {NoStop}%
\bibitem [{\citenamefont {Blatter}\ \emph {et~al.}(1994)\citenamefont
  {Blatter}, \citenamefont {Feigel'man}, \citenamefont {Geshkenbein},
  \citenamefont {Larkin},\ and\ \citenamefont {Vinokur}}]{Blatter94}%
  \BibitemOpen
  \bibfield  {author} {\bibinfo {author} {\bibfnamefont {G.}~\bibnamefont
  {Blatter}}, \bibinfo {author} {\bibfnamefont {M.~V.}\ \bibnamefont
  {Feigel'man}}, \bibinfo {author} {\bibfnamefont {V.~B.}\ \bibnamefont
  {Geshkenbein}}, \bibinfo {author} {\bibfnamefont {A.~I.}\ \bibnamefont
  {Larkin}}, \ and\ \bibinfo {author} {\bibfnamefont {V.~M.}\ \bibnamefont
  {Vinokur}},\ }\bibfield  {title} {\enquote {\bibinfo {title} {Vortices in
  high-temperature superconductors},}\ }\href {\doibase
  10.1103/RevModPhys.66.1125} {\bibfield  {journal} {\bibinfo  {journal} {Rev.
  Mod. Phys.}\ }\textbf {\bibinfo {volume} {66}},\ \bibinfo {pages}
  {1125--1388} (\bibinfo {year} {1994})}\BibitemShut {NoStop}%
\bibitem [{\citenamefont {Reichhardt}\ and\ \citenamefont
  {Olson}(2002)}]{Reichhardt02}%
  \BibitemOpen
  \bibfield  {author} {\bibinfo {author} {\bibfnamefont {C.}~\bibnamefont
  {Reichhardt}}\ and\ \bibinfo {author} {\bibfnamefont {C.~J.}\ \bibnamefont
  {Olson}},\ }\bibfield  {title} {\enquote {\bibinfo {title} {Colloidal
  dynamics on disordered substrates},}\ }\href {\doibase
  10.1103/PhysRevLett.89.078301} {\bibfield  {journal} {\bibinfo  {journal}
  {Phys. Rev. Lett.}\ }\textbf {\bibinfo {volume} {89}},\ \bibinfo {pages}
  {078301} (\bibinfo {year} {2002})}\BibitemShut {NoStop}%
\bibitem [{\citenamefont {Pertsinidis}\ and\ \citenamefont
  {Ling}(2008)}]{Pertsinidis08}%
  \BibitemOpen
  \bibfield  {author} {\bibinfo {author} {\bibfnamefont {A.}~\bibnamefont
  {Pertsinidis}}\ and\ \bibinfo {author} {\bibfnamefont {X.~S.}\ \bibnamefont
  {Ling}},\ }\bibfield  {title} {\enquote {\bibinfo {title} {Statics and
  dynamics of {2D} colloidal crystals in a random pinning potential},}\ }\href
  {\doibase 10.1103/PhysRevLett.100.028303} {\bibfield  {journal} {\bibinfo
  {journal} {Phys. Rev. Lett.}\ }\textbf {\bibinfo {volume} {100}},\ \bibinfo
  {pages} {028303} (\bibinfo {year} {2008})}\BibitemShut {NoStop}%
\bibitem [{\citenamefont {Tierno}(2012)}]{Tierno12a}%
  \BibitemOpen
  \bibfield  {author} {\bibinfo {author} {\bibfnamefont {P.}~\bibnamefont
  {Tierno}},\ }\bibfield  {title} {\enquote {\bibinfo {title} {Depinning and
  collective dynamics of magnetically driven colloidal monolayers},}\ }\href
  {\doibase 10.1103/PhysRevLett.109.198304} {\bibfield  {journal} {\bibinfo
  {journal} {Phys. Rev. Lett.}\ }\textbf {\bibinfo {volume} {109}},\ \bibinfo
  {pages} {198304} (\bibinfo {year} {2012})}\BibitemShut {NoStop}%
\bibitem [{\citenamefont {Vanossi}\ \emph {et~al.}(2013)\citenamefont
  {Vanossi}, \citenamefont {Manini}, \citenamefont {Urbakh}, \citenamefont
  {Zapperi},\ and\ \citenamefont {Tosatti}}]{Vanossi13}%
  \BibitemOpen
  \bibfield  {author} {\bibinfo {author} {\bibfnamefont {A.}~\bibnamefont
  {Vanossi}}, \bibinfo {author} {\bibfnamefont {N.}~\bibnamefont {Manini}},
  \bibinfo {author} {\bibfnamefont {M.}~\bibnamefont {Urbakh}}, \bibinfo
  {author} {\bibfnamefont {S.}~\bibnamefont {Zapperi}}, \ and\ \bibinfo
  {author} {\bibfnamefont {E.}~\bibnamefont {Tosatti}},\ }\bibfield  {title}
  {\enquote {\bibinfo {title} {Colloquium: Modeling friction: From nanoscale to
  mesoscale},}\ }\href {\doibase 10.1103/RevModPhys.85.529} {\bibfield
  {journal} {\bibinfo  {journal} {Rev. Mod. Phys.}\ }\textbf {\bibinfo {volume}
  {85}},\ \bibinfo {pages} {529--552} (\bibinfo {year} {2013})}\BibitemShut
  {NoStop}%
\bibitem [{\citenamefont {Fisher}(1998)}]{Fisher98}%
  \BibitemOpen
  \bibfield  {author} {\bibinfo {author} {\bibfnamefont {D.~S.}\ \bibnamefont
  {Fisher}},\ }\bibfield  {title} {\enquote {\bibinfo {title} {Collective
  transport in random media: from superconductors to earthquakes},}\ }\href
  {\doibase 10.1016/S0370-1573(98)00008-8} {\bibfield  {journal} {\bibinfo
  {journal} {Phys. Rep.}\ }\textbf {\bibinfo {volume} {301}},\ \bibinfo {pages}
  {113--150} (\bibinfo {year} {1998})}\BibitemShut {NoStop}%
\bibitem [{\citenamefont {Reichhardt}\ and\ \citenamefont
  {Reichhardt}(2017{\natexlab{a}})}]{Reichhardt17}%
  \BibitemOpen
  \bibfield  {author} {\bibinfo {author} {\bibfnamefont {C.}~\bibnamefont
  {Reichhardt}}\ and\ \bibinfo {author} {\bibfnamefont {C.~J.~Olson}\
  \bibnamefont {Reichhardt}},\ }\bibfield  {title} {\enquote {\bibinfo {title}
  {Depinning and nonequilibrium dynamic phases of particle assemblies driven
  over random and ordered substrates: a review},}\ }\href {\doibase
  10.1088/1361-6633/80/2/026501} {\bibfield  {journal} {\bibinfo  {journal}
  {Rep. Prog. Phys.}\ }\textbf {\bibinfo {volume} {80}},\ \bibinfo {pages}
  {026501} (\bibinfo {year} {2017}{\natexlab{a}})}\BibitemShut {NoStop}%
\bibitem [{\citenamefont {Pardo}\ \emph {et~al.}(1998)\citenamefont {Pardo},
  \citenamefont {de~la Cruz}, \citenamefont {Gammel}, \citenamefont {Bucher},\
  and\ \citenamefont {Bishop}}]{Pardo98}%
  \BibitemOpen
  \bibfield  {author} {\bibinfo {author} {\bibfnamefont {F.}~\bibnamefont
  {Pardo}}, \bibinfo {author} {\bibfnamefont {F.}~\bibnamefont {de~la Cruz}},
  \bibinfo {author} {\bibfnamefont {P.~L.}\ \bibnamefont {Gammel}}, \bibinfo
  {author} {\bibfnamefont {E.}~\bibnamefont {Bucher}}, \ and\ \bibinfo {author}
  {\bibfnamefont {D.~J.}\ \bibnamefont {Bishop}},\ }\bibfield  {title}
  {\enquote {\bibinfo {title} {Observation of smectic and moving-{Bragg}-glass
  phases in flowing vortex lattices},}\ }\href {\doibase 10.1038/24581}
  {\bibfield  {journal} {\bibinfo  {journal} {Nature}\ }\textbf {\bibinfo
  {volume} {396}},\ \bibinfo {pages} {348--350} (\bibinfo {year}
  {1998})}\BibitemShut {NoStop}%
\bibitem [{\citenamefont {Olson}\ \emph {et~al.}(1998)\citenamefont {Olson},
  \citenamefont {Reichhardt},\ and\ \citenamefont {Nori}}]{Olson98a}%
  \BibitemOpen
  \bibfield  {author} {\bibinfo {author} {\bibfnamefont {C.~J.}\ \bibnamefont
  {Olson}}, \bibinfo {author} {\bibfnamefont {C.}~\bibnamefont {Reichhardt}}, \
  and\ \bibinfo {author} {\bibfnamefont {F.}~\bibnamefont {Nori}},\ }\bibfield
  {title} {\enquote {\bibinfo {title} {Nonequilibrium dynamic phase diagram for
  vortex lattices},}\ }\href {\doibase 10.1103/PhysRevLett.81.3757} {\bibfield
  {journal} {\bibinfo  {journal} {Phys. Rev. Lett.}\ }\textbf {\bibinfo
  {volume} {81}},\ \bibinfo {pages} {3757--3760} (\bibinfo {year}
  {1998})}\BibitemShut {NoStop}%
\bibitem [{\citenamefont {Dzubiella}\ \emph {et~al.}(2002)\citenamefont
  {Dzubiella}, \citenamefont {Hoffmann},\ and\ \citenamefont
  {L\"owen}}]{Dzubiella02a}%
  \BibitemOpen
  \bibfield  {author} {\bibinfo {author} {\bibfnamefont {J.}~\bibnamefont
  {Dzubiella}}, \bibinfo {author} {\bibfnamefont {G.~P.}\ \bibnamefont
  {Hoffmann}}, \ and\ \bibinfo {author} {\bibfnamefont {H.}~\bibnamefont
  {L\"owen}},\ }\bibfield  {title} {\enquote {\bibinfo {title} {Lane formation
  in colloidal mixtures driven by an external field},}\ }\href {\doibase
  10.1103/PhysRevE.65.021402} {\bibfield  {journal} {\bibinfo  {journal} {Phys.
  Rev. E}\ }\textbf {\bibinfo {volume} {65}},\ \bibinfo {pages} {021402}
  (\bibinfo {year} {2002})}\BibitemShut {NoStop}%
\bibitem [{\citenamefont {Vissers}\ \emph {et~al.}(2011)\citenamefont
  {Vissers}, \citenamefont {van Blaaderen},\ and\ \citenamefont
  {Imhof}}]{Vissers11a}%
  \BibitemOpen
  \bibfield  {author} {\bibinfo {author} {\bibfnamefont {T.}~\bibnamefont
  {Vissers}}, \bibinfo {author} {\bibfnamefont {A.}~\bibnamefont {van
  Blaaderen}}, \ and\ \bibinfo {author} {\bibfnamefont {A.}~\bibnamefont
  {Imhof}},\ }\bibfield  {title} {\enquote {\bibinfo {title} {Band formation in
  mixtures of oppositely charged colloids driven by an ac electric field},}\
  }\href {\doibase 10.1103/PhysRevLett.106.228303} {\bibfield  {journal}
  {\bibinfo  {journal} {Phys. Rev. Lett.}\ }\textbf {\bibinfo {volume} {106}},\
  \bibinfo {pages} {228303} (\bibinfo {year} {2011})}\BibitemShut {NoStop}%
\bibitem [{\citenamefont {Reichhardt}\ and\ \citenamefont
  {Reichhardt}(2018)}]{Reichhardt18}%
  \BibitemOpen
  \bibfield  {author} {\bibinfo {author} {\bibfnamefont {C.}~\bibnamefont
  {Reichhardt}}\ and\ \bibinfo {author} {\bibfnamefont {C.~J.~O.}\ \bibnamefont
  {Reichhardt}},\ }\bibfield  {title} {\enquote {\bibinfo {title} {Velocity
  force curves, laning, and jamming for oppositely driven disk systems},}\
  }\href {\doibase 10.1039/c7sm02162c} {\bibfield  {journal} {\bibinfo
  {journal} {Soft Matter}\ }\textbf {\bibinfo {volume} {14}},\ \bibinfo {pages}
  {490--498} (\bibinfo {year} {2018})}\BibitemShut {NoStop}%
\bibitem [{\citenamefont {Marchetti}\ \emph {et~al.}(2013)\citenamefont
  {Marchetti}, \citenamefont {Joanny}, \citenamefont {Ramaswamy}, \citenamefont
  {Liverpool}, \citenamefont {Prost}, \citenamefont {Rao},\ and\ \citenamefont
  {Simha}}]{Marchetti13}%
  \BibitemOpen
  \bibfield  {author} {\bibinfo {author} {\bibfnamefont {M.~C.}\ \bibnamefont
  {Marchetti}}, \bibinfo {author} {\bibfnamefont {J.~F.}\ \bibnamefont
  {Joanny}}, \bibinfo {author} {\bibfnamefont {S.}~\bibnamefont {Ramaswamy}},
  \bibinfo {author} {\bibfnamefont {T.~B.}\ \bibnamefont {Liverpool}}, \bibinfo
  {author} {\bibfnamefont {J.}~\bibnamefont {Prost}}, \bibinfo {author}
  {\bibfnamefont {M.}~\bibnamefont {Rao}}, \ and\ \bibinfo {author}
  {\bibfnamefont {R.~A.}\ \bibnamefont {Simha}},\ }\bibfield  {title} {\enquote
  {\bibinfo {title} {Hydrodynamics of soft active matter},}\ }\href {\doibase
  10.1103/RevModPhys.85.1143} {\bibfield  {journal} {\bibinfo  {journal} {Rev.
  Mod. Phys.}\ }\textbf {\bibinfo {volume} {85}},\ \bibinfo {pages}
  {1143--1189} (\bibinfo {year} {2013})}\BibitemShut {NoStop}%
\bibitem [{\citenamefont {Bechinger}\ \emph {et~al.}(2016)\citenamefont
  {Bechinger}, \citenamefont {Di~Leonardo}, \citenamefont {L\"owen},
  \citenamefont {Reichhardt}, \citenamefont {Volpe},\ and\ \citenamefont
  {Volpe}}]{Bechinger16}%
  \BibitemOpen
  \bibfield  {author} {\bibinfo {author} {\bibfnamefont {C.}~\bibnamefont
  {Bechinger}}, \bibinfo {author} {\bibfnamefont {R.}~\bibnamefont
  {Di~Leonardo}}, \bibinfo {author} {\bibfnamefont {H.}~\bibnamefont
  {L\"owen}}, \bibinfo {author} {\bibfnamefont {C.}~\bibnamefont {Reichhardt}},
  \bibinfo {author} {\bibfnamefont {G.}~\bibnamefont {Volpe}}, \ and\ \bibinfo
  {author} {\bibfnamefont {G.}~\bibnamefont {Volpe}},\ }\bibfield  {title}
  {\enquote {\bibinfo {title} {Active particles in complex and crowded
  environments},}\ }\href {\doibase 10.1103/RevModPhys.88.045006} {\bibfield
  {journal} {\bibinfo  {journal} {Rev. Mod. Phys.}\ }\textbf {\bibinfo {volume}
  {88}},\ \bibinfo {pages} {045006} (\bibinfo {year} {2016})}\BibitemShut
  {NoStop}%
\bibitem [{\citenamefont {Fily}\ and\ \citenamefont
  {Marchetti}(2012)}]{Fily12}%
  \BibitemOpen
  \bibfield  {author} {\bibinfo {author} {\bibfnamefont {Y.}~\bibnamefont
  {Fily}}\ and\ \bibinfo {author} {\bibfnamefont {M.~C.}\ \bibnamefont
  {Marchetti}},\ }\bibfield  {title} {\enquote {\bibinfo {title} {Athermal
  phase separation of self-propelled particles with no alignment},}\ }\href
  {\doibase 10.1103/PhysRevLett.108.235702} {\bibfield  {journal} {\bibinfo
  {journal} {Phys. Rev. Lett.}\ }\textbf {\bibinfo {volume} {108}},\ \bibinfo
  {pages} {235702} (\bibinfo {year} {2012})}\BibitemShut {NoStop}%
\bibitem [{\citenamefont {Redner}\ \emph {et~al.}(2013)\citenamefont {Redner},
  \citenamefont {Hagan},\ and\ \citenamefont {Baskaran}}]{Redner13}%
  \BibitemOpen
  \bibfield  {author} {\bibinfo {author} {\bibfnamefont {G.~S.}\ \bibnamefont
  {Redner}}, \bibinfo {author} {\bibfnamefont {M.~F.}\ \bibnamefont {Hagan}}, \
  and\ \bibinfo {author} {\bibfnamefont {A.}~\bibnamefont {Baskaran}},\
  }\bibfield  {title} {\enquote {\bibinfo {title} {Structure and dynamics of a
  phase-separating active colloidal fluid},}\ }\href {\doibase
  10.1103/PhysRevLett.110.055701} {\bibfield  {journal} {\bibinfo  {journal}
  {Phys. Rev. Lett.}\ }\textbf {\bibinfo {volume} {110}},\ \bibinfo {pages}
  {055701} (\bibinfo {year} {2013})}\BibitemShut {NoStop}%
\bibitem [{\citenamefont {Palacci}\ \emph {et~al.}(2013)\citenamefont
  {Palacci}, \citenamefont {Sacanna}, \citenamefont {Steinberg}, \citenamefont
  {Pine},\ and\ \citenamefont {Chaikin}}]{Palacci13}%
  \BibitemOpen
  \bibfield  {author} {\bibinfo {author} {\bibfnamefont {J.}~\bibnamefont
  {Palacci}}, \bibinfo {author} {\bibfnamefont {S.}~\bibnamefont {Sacanna}},
  \bibinfo {author} {\bibfnamefont {A.~P.}\ \bibnamefont {Steinberg}}, \bibinfo
  {author} {\bibfnamefont {D.~J.}\ \bibnamefont {Pine}}, \ and\ \bibinfo
  {author} {\bibfnamefont {P.~M.}\ \bibnamefont {Chaikin}},\ }\bibfield
  {title} {\enquote {\bibinfo {title} {Living crystals of light-activated
  colloidal surfers},}\ }\href {\doibase 10.1126/science.1230020} {\bibfield
  {journal} {\bibinfo  {journal} {Science}\ }\textbf {\bibinfo {volume}
  {339}},\ \bibinfo {pages} {936--940} (\bibinfo {year} {2013})}\BibitemShut
  {NoStop}%
\bibitem [{\citenamefont {Buttinoni}\ \emph {et~al.}(2013)\citenamefont
  {Buttinoni}, \citenamefont {Bialk\'e}, \citenamefont {K\"ummel},
  \citenamefont {L\"owen}, \citenamefont {Bechinger},\ and\ \citenamefont
  {Speck}}]{Buttinoni13}%
  \BibitemOpen
  \bibfield  {author} {\bibinfo {author} {\bibfnamefont {I.}~\bibnamefont
  {Buttinoni}}, \bibinfo {author} {\bibfnamefont {J.}~\bibnamefont {Bialk\'e}},
  \bibinfo {author} {\bibfnamefont {F.}~\bibnamefont {K\"ummel}}, \bibinfo
  {author} {\bibfnamefont {H.}~\bibnamefont {L\"owen}}, \bibinfo {author}
  {\bibfnamefont {C.}~\bibnamefont {Bechinger}}, \ and\ \bibinfo {author}
  {\bibfnamefont {T.}~\bibnamefont {Speck}},\ }\bibfield  {title} {\enquote
  {\bibinfo {title} {Dynamical clustering and phase separation in suspensions
  of self-propelled colloidal particles},}\ }\href {\doibase
  10.1103/PhysRevLett.110.238301} {\bibfield  {journal} {\bibinfo  {journal}
  {Phys. Rev. Lett.}\ }\textbf {\bibinfo {volume} {110}},\ \bibinfo {pages}
  {238301} (\bibinfo {year} {2013})}\BibitemShut {NoStop}%
\bibitem [{\citenamefont {Cates}\ and\ \citenamefont
  {Tailleur}(2015)}]{Cates15}%
  \BibitemOpen
  \bibfield  {author} {\bibinfo {author} {\bibfnamefont {M.~E.}\ \bibnamefont
  {Cates}}\ and\ \bibinfo {author} {\bibfnamefont {J.}~\bibnamefont
  {Tailleur}},\ }\bibfield  {title} {\enquote {\bibinfo {title}
  {Motility-induced phase separation},}\ }\href {\doibase
  10.1146/annurev-conmatphys-031214-014710} {\bibfield  {journal} {\bibinfo
  {journal} {Annual Review of Condensed Matter Physics}\ }\textbf {\bibinfo
  {volume} {6}},\ \bibinfo {pages} {219--244} (\bibinfo {year}
  {2015})}\BibitemShut {NoStop}%
\bibitem [{\citenamefont {Reichhardt}\ and\ \citenamefont
  {Reichhardt}(2015)}]{Reichhardt15}%
  \BibitemOpen
  \bibfield  {author} {\bibinfo {author} {\bibfnamefont {C.}~\bibnamefont
  {Reichhardt}}\ and\ \bibinfo {author} {\bibfnamefont {C.~J.~Olson}\
  \bibnamefont {Reichhardt}},\ }\bibfield  {title} {\enquote {\bibinfo {title}
  {Active microrheology in active matter systems: Mobility, intermittency, and
  avalanches},}\ }\href {\doibase 10.1103/PhysRevE.91.032313} {\bibfield
  {journal} {\bibinfo  {journal} {Phys. Rev. E}\ }\textbf {\bibinfo {volume}
  {91}},\ \bibinfo {pages} {032313} (\bibinfo {year} {2015})}\BibitemShut
  {NoStop}%
\bibitem [{\citenamefont {S\'andor}\ \emph {et~al.}(2017)\citenamefont
  {S\'andor}, \citenamefont {Lib\'al}, \citenamefont {Reichhardt},\ and\
  \citenamefont {Olson~Reichhardt}}]{Sandor17a}%
  \BibitemOpen
  \bibfield  {author} {\bibinfo {author} {\bibfnamefont {Cs.}\ \bibnamefont
  {S\'andor}}, \bibinfo {author} {\bibfnamefont {A.}~\bibnamefont {Lib\'al}},
  \bibinfo {author} {\bibfnamefont {C.}~\bibnamefont {Reichhardt}}, \ and\
  \bibinfo {author} {\bibfnamefont {C.~J.}\ \bibnamefont {Olson~Reichhardt}},\
  }\bibfield  {title} {\enquote {\bibinfo {title} {Dynamic phases of active
  matter systems with quenched disorder},}\ }\href {\doibase
  10.1103/PhysRevE.95.032606} {\bibfield  {journal} {\bibinfo  {journal} {Phys.
  Rev. E}\ }\textbf {\bibinfo {volume} {95}},\ \bibinfo {pages} {032606}
  (\bibinfo {year} {2017})}\BibitemShut {NoStop}%
\bibitem [{\citenamefont {Vicsek}\ and\ \citenamefont
  {Zaferis}(2012)}]{Vicsek12}%
  \BibitemOpen
  \bibfield  {author} {\bibinfo {author} {\bibfnamefont {T.}~\bibnamefont
  {Vicsek}}\ and\ \bibinfo {author} {\bibfnamefont {A.}~\bibnamefont
  {Zaferis}},\ }\bibfield  {title} {\enquote {\bibinfo {title} {Collective
  motion},}\ }\href {\doibase 10.1016/j.physrep.2012.03.004} {\bibfield
  {journal} {\bibinfo  {journal} {Phys. Rep.}\ }\textbf {\bibinfo {volume}
  {517}},\ \bibinfo {pages} {71} (\bibinfo {year} {2012})}\BibitemShut
  {NoStop}%
\bibitem [{\citenamefont {Morin}\ \emph {et~al.}(2017)\citenamefont {Morin},
  \citenamefont {Desreumaux}, \citenamefont {Caussin},\ and\ \citenamefont
  {Bartolo}}]{Morin17}%
  \BibitemOpen
  \bibfield  {author} {\bibinfo {author} {\bibfnamefont {A.}~\bibnamefont
  {Morin}}, \bibinfo {author} {\bibfnamefont {N.}~\bibnamefont {Desreumaux}},
  \bibinfo {author} {\bibfnamefont {J.-B.}\ \bibnamefont {Caussin}}, \ and\
  \bibinfo {author} {\bibfnamefont {D.}~\bibnamefont {Bartolo}},\ }\bibfield
  {title} {\enquote {\bibinfo {title} {Distortion and destruction of colloidal
  flocks in disordered environments},}\ }\href {\doibase 10.1038/nphys3903}
  {\bibfield  {journal} {\bibinfo  {journal} {Nature Phys.}\ }\textbf {\bibinfo
  {volume} {13}},\ \bibinfo {pages} {63--67} (\bibinfo {year}
  {2017})}\BibitemShut {NoStop}%
\bibitem [{\citenamefont {Wang}\ \emph {et~al.}(2021)\citenamefont {Wang},
  \citenamefont {Phan}, \citenamefont {Li}, \citenamefont {Wombacher},
  \citenamefont {Qu}, \citenamefont {Peng}, \citenamefont {Chen}, \citenamefont
  {Goldman}, \citenamefont {Levin}, \citenamefont {Austin},\ and\ \citenamefont
  {Liu}}]{Wang21}%
  \BibitemOpen
  \bibfield  {author} {\bibinfo {author} {\bibfnamefont {G.}~\bibnamefont
  {Wang}}, \bibinfo {author} {\bibfnamefont {T.~V.}\ \bibnamefont {Phan}},
  \bibinfo {author} {\bibfnamefont {S.}~\bibnamefont {Li}}, \bibinfo {author}
  {\bibfnamefont {M.}~\bibnamefont {Wombacher}}, \bibinfo {author}
  {\bibfnamefont {J.}~\bibnamefont {Qu}}, \bibinfo {author} {\bibfnamefont
  {Y.}~\bibnamefont {Peng}}, \bibinfo {author} {\bibfnamefont {G.}~\bibnamefont
  {Chen}}, \bibinfo {author} {\bibfnamefont {D.~I.}\ \bibnamefont {Goldman}},
  \bibinfo {author} {\bibfnamefont {S.~A.}\ \bibnamefont {Levin}}, \bibinfo
  {author} {\bibfnamefont {R.~H.}\ \bibnamefont {Austin}}, \ and\ \bibinfo
  {author} {\bibfnamefont {L.}~\bibnamefont {Liu}},\ }\bibfield  {title}
  {\enquote {\bibinfo {title} {Emergent field-driven robot swarm states},}\
  }\href {\doibase 10.1103/PhysRevLett.126.108002} {\bibfield  {journal}
  {\bibinfo  {journal} {Phys. Rev. Lett.}\ }\textbf {\bibinfo {volume} {126}},\
  \bibinfo {pages} {108002} (\bibinfo {year} {2021})}\BibitemShut {NoStop}%
\bibitem [{\citenamefont {Wang}\ \emph {et~al.}(2022)\citenamefont {Wang},
  \citenamefont {Phan}, \citenamefont {Li}, \citenamefont {Wang}, \citenamefont
  {Peng}, \citenamefont {Chen}, \citenamefont {Qu}, \citenamefont {Goldman},
  \citenamefont {Levin}, \citenamefont {Pienta}, \citenamefont {Amend},
  \citenamefont {Austin},\ and\ \citenamefont {Liu}}]{Wang22}%
  \BibitemOpen
  \bibfield  {author} {\bibinfo {author} {\bibfnamefont {G.}~\bibnamefont
  {Wang}}, \bibinfo {author} {\bibfnamefont {T.~V.}\ \bibnamefont {Phan}},
  \bibinfo {author} {\bibfnamefont {S.}~\bibnamefont {Li}}, \bibinfo {author}
  {\bibfnamefont {J.}~\bibnamefont {Wang}}, \bibinfo {author} {\bibfnamefont
  {Y.}~\bibnamefont {Peng}}, \bibinfo {author} {\bibfnamefont {G.}~\bibnamefont
  {Chen}}, \bibinfo {author} {\bibfnamefont {J.}~\bibnamefont {Qu}}, \bibinfo
  {author} {\bibfnamefont {D.~I.}\ \bibnamefont {Goldman}}, \bibinfo {author}
  {\bibfnamefont {S.~A.}\ \bibnamefont {Levin}}, \bibinfo {author}
  {\bibfnamefont {K.}~\bibnamefont {Pienta}}, \bibinfo {author} {\bibfnamefont
  {S.}~\bibnamefont {Amend}}, \bibinfo {author} {\bibfnamefont {R.~H.}\
  \bibnamefont {Austin}}, \ and\ \bibinfo {author} {\bibfnamefont
  {L.}~\bibnamefont {Liu}},\ }\bibfield  {title} {\enquote {\bibinfo {title}
  {Robots as models of evolving systems},}\ }\href {\doibase
  10.1073/pnas.2120019119} {\bibfield  {journal} {\bibinfo  {journal} {Proc.
  Natl. Acad. Sci. (USA)}\ }\textbf {\bibinfo {volume} {119}},\ \bibinfo
  {pages} {e2120019119} (\bibinfo {year} {2022})}\BibitemShut {NoStop}%
\bibitem [{\citenamefont {Varga}\ \emph {et~al.}(2022)\citenamefont {Varga},
  \citenamefont {Lib\'al}, \citenamefont {Reichhardt},\ and\ \citenamefont
  {Reichhardt}}]{Varga22}%
  \BibitemOpen
  \bibfield  {author} {\bibinfo {author} {\bibfnamefont {L.}~\bibnamefont
  {Varga}}, \bibinfo {author} {\bibfnamefont {A.}~\bibnamefont {Lib\'al}},
  \bibinfo {author} {\bibfnamefont {C.~J.~O.}\ \bibnamefont {Reichhardt}}, \
  and\ \bibinfo {author} {\bibfnamefont {C.}~\bibnamefont {Reichhardt}},\
  }\bibfield  {title} {\enquote {\bibinfo {title} {Active regimes for particles
  on resource landscapes},}\ }\href {\doibase 10.1103/PhysRevResearch.4.013061}
  {\bibfield  {journal} {\bibinfo  {journal} {Phys. Rev. Research}\ }\textbf
  {\bibinfo {volume} {4}},\ \bibinfo {pages} {013061} (\bibinfo {year}
  {2022})}\BibitemShut {NoStop}%
\bibitem [{\citenamefont {Reichhardt}\ and\ \citenamefont
  {Reichhardt}(2014{\natexlab{a}})}]{Reichhardt14}%
  \BibitemOpen
  \bibfield  {author} {\bibinfo {author} {\bibfnamefont {C.}~\bibnamefont
  {Reichhardt}}\ and\ \bibinfo {author} {\bibfnamefont {C.~J.~Olson}\
  \bibnamefont {Reichhardt}},\ }\bibfield  {title} {\enquote {\bibinfo {title}
  {Aspects of jamming in two-dimensional athermal frictionless systems},}\
  }\href {\doibase 10.1039/c3sm53154f} {\bibfield  {journal} {\bibinfo
  {journal} {Soft Matter}\ }\textbf {\bibinfo {volume} {10}},\ \bibinfo {pages}
  {2932--2944} (\bibinfo {year} {2014}{\natexlab{a}})}\BibitemShut {NoStop}%
\bibitem [{\citenamefont {Zykov}\ and\ \citenamefont
  {Bodenschatz}(2018)}]{Zykov18}%
  \BibitemOpen
  \bibfield  {author} {\bibinfo {author} {\bibfnamefont {V.~S.}\ \bibnamefont
  {Zykov}}\ and\ \bibinfo {author} {\bibfnamefont {E.}~\bibnamefont
  {Bodenschatz}},\ }\bibfield  {title} {\enquote {\bibinfo {title} {Wave
  propagation in inhomogeneous excitable media},}\ }\href {\doibase
  10.1146/annurev-conmatphys-033117-054300} {\bibfield  {journal} {\bibinfo
  {journal} {Ann. Rev. Condens. Matter Phys.}\ }\textbf {\bibinfo {volume}
  {9}},\ \bibinfo {pages} {435} (\bibinfo {year} {2018})}\BibitemShut {NoStop}%
\bibitem [{\citenamefont {Seul}\ and\ \citenamefont {Andelman}(1995)}]{Seul95}%
  \BibitemOpen
  \bibfield  {author} {\bibinfo {author} {\bibfnamefont {M.}~\bibnamefont
  {Seul}}\ and\ \bibinfo {author} {\bibfnamefont {D.}~\bibnamefont
  {Andelman}},\ }\bibfield  {title} {\enquote {\bibinfo {title} {Domain shapes
  and patterns - the phenomenology of modulated phases},}\ }\href {\doibase
  10.1126/science.267.5197.476} {\bibfield  {journal} {\bibinfo  {journal}
  {Science}\ }\textbf {\bibinfo {volume} {267}},\ \bibinfo {pages} {476--483}
  (\bibinfo {year} {1995})}\BibitemShut {NoStop}%
\bibitem [{\citenamefont {Malescio}\ and\ \citenamefont
  {Pellicane}(2003)}]{Malescio03}%
  \BibitemOpen
  \bibfield  {author} {\bibinfo {author} {\bibfnamefont {G.}~\bibnamefont
  {Malescio}}\ and\ \bibinfo {author} {\bibfnamefont {G.}~\bibnamefont
  {Pellicane}},\ }\bibfield  {title} {\enquote {\bibinfo {title} {Stripe phases
  from isotropic repulsive interactions},}\ }\href {\doibase 10.1038/nmat820}
  {\bibfield  {journal} {\bibinfo  {journal} {Nature Mater.}\ }\textbf
  {\bibinfo {volume} {2}},\ \bibinfo {pages} {97--100} (\bibinfo {year}
  {2003})}\BibitemShut {NoStop}%
\bibitem [{\citenamefont {Olson~Reichhardt}\ \emph {et~al.}(2010)\citenamefont
  {Olson~Reichhardt}, \citenamefont {Reichhardt},\ and\ \citenamefont
  {Bishop}}]{Reichhardt10}%
  \BibitemOpen
  \bibfield  {author} {\bibinfo {author} {\bibfnamefont {C.~J.}\ \bibnamefont
  {Olson~Reichhardt}}, \bibinfo {author} {\bibfnamefont {C.}~\bibnamefont
  {Reichhardt}}, \ and\ \bibinfo {author} {\bibfnamefont {A.~R.}\ \bibnamefont
  {Bishop}},\ }\bibfield  {title} {\enquote {\bibinfo {title} {Structural
  transitions, melting, and intermediate phases for stripe- and clump-forming
  systems},}\ }\href {\doibase 10.1103/PhysRevE.82.041502} {\bibfield
  {journal} {\bibinfo  {journal} {Phys. Rev. E}\ }\textbf {\bibinfo {volume}
  {82}},\ \bibinfo {pages} {041502} (\bibinfo {year} {2010})}\BibitemShut
  {NoStop}%
\bibitem [{\citenamefont {Neto}\ and\ \citenamefont {Silva}(2022)}]{Neto22}%
  \BibitemOpen
  \bibfield  {author} {\bibinfo {author} {\bibfnamefont {J.~F.}\ \bibnamefont
  {Neto}}\ and\ \bibinfo {author} {\bibfnamefont {C.~C. de~Souza}\ \bibnamefont
  {Silva}},\ }\bibfield  {title} {\enquote {\bibinfo {title} {Mesoscale phase
  separation of skyrmion-vortex matter in chiral-magnet--superconductor
  heterostructures},}\ }\href {\doibase 10.1103/PhysRevLett.128.057001}
  {\bibfield  {journal} {\bibinfo  {journal} {Phys. Rev. Lett.}\ }\textbf
  {\bibinfo {volume} {128}},\ \bibinfo {pages} {057001} (\bibinfo {year}
  {2022})}\BibitemShut {NoStop}%
\bibitem [{\citenamefont {Forg\'acs}\ \emph {et~al.}(2021)\citenamefont
  {Forg\'acs}, \citenamefont {Lib\'al}, \citenamefont {Reichhardt},\ and\
  \citenamefont {Reichhardt}}]{Forgacs21}%
  \BibitemOpen
  \bibfield  {author} {\bibinfo {author} {\bibfnamefont {P.}~\bibnamefont
  {Forg\'acs}}, \bibinfo {author} {\bibfnamefont {A.}~\bibnamefont {Lib\'al}},
  \bibinfo {author} {\bibfnamefont {C.}~\bibnamefont {Reichhardt}}, \ and\
  \bibinfo {author} {\bibfnamefont {C.~J.~O.}\ \bibnamefont {Reichhardt}},\
  }\bibfield  {title} {\enquote {\bibinfo {title} {Active matter shepherding
  and clustering in inhomogeneous environments},}\ }\href {\doibase
  10.1103/PhysRevE.104.044613} {\bibfield  {journal} {\bibinfo  {journal}
  {Phys. Rev. E}\ }\textbf {\bibinfo {volume} {104}},\ \bibinfo {pages}
  {044613} (\bibinfo {year} {2021})}\BibitemShut {NoStop}%
\bibitem [{\citenamefont {Ni}\ \emph {et~al.}(2014)\citenamefont {Ni},
  \citenamefont {Cohen~Stuart}, \citenamefont {Dijkstra},\ and\ \citenamefont
  {Bolhuis}}]{Ni14}%
  \BibitemOpen
  \bibfield  {author} {\bibinfo {author} {\bibfnamefont {R.}~\bibnamefont
  {Ni}}, \bibinfo {author} {\bibfnamefont {M.~A.}\ \bibnamefont
  {Cohen~Stuart}}, \bibinfo {author} {\bibfnamefont {M.}~\bibnamefont
  {Dijkstra}}, \ and\ \bibinfo {author} {\bibfnamefont {P.~G.}\ \bibnamefont
  {Bolhuis}},\ }\bibfield  {title} {\enquote {\bibinfo {title} {Crystallizing
  hard-sphere glasses by doping with active particles},}\ }\href {\doibase
  10.1039/C4SM01015A} {\bibfield  {journal} {\bibinfo  {journal} {Soft Matter}\
  }\textbf {\bibinfo {volume} {10}},\ \bibinfo {pages} {6609} (\bibinfo {year}
  {2014})}\BibitemShut {NoStop}%
\bibitem [{\citenamefont {K{\" u}mmel}\ \emph {et~al.}(2015)\citenamefont {K{\"
  u}mmel}, \citenamefont {Shabestari}, \citenamefont {Lozano}, \citenamefont
  {Volpe},\ and\ \citenamefont {Bechinger}}]{Kummel15}%
  \BibitemOpen
  \bibfield  {author} {\bibinfo {author} {\bibfnamefont {F.}~\bibnamefont {K{\"
  u}mmel}}, \bibinfo {author} {\bibfnamefont {P.}~\bibnamefont {Shabestari}},
  \bibinfo {author} {\bibfnamefont {C.}~\bibnamefont {Lozano}}, \bibinfo
  {author} {\bibfnamefont {G.}~\bibnamefont {Volpe}}, \ and\ \bibinfo {author}
  {\bibfnamefont {C.}~\bibnamefont {Bechinger}},\ }\bibfield  {title} {\enquote
  {\bibinfo {title} {Formation, compression and surface melting of colloidal
  clusters by active particles},}\ }\href {\doibase 10.1039/c5sm00827a}
  {\bibfield  {journal} {\bibinfo  {journal} {Soft Matter}\ }\textbf {\bibinfo
  {volume} {11}},\ \bibinfo {pages} {6187--6191} (\bibinfo {year}
  {2015})}\BibitemShut {NoStop}%
\bibitem [{\citenamefont {Ramananarivo}\ \emph {et~al.}(2019)\citenamefont
  {Ramananarivo}, \citenamefont {Ducrot},\ and\ \citenamefont
  {Palacci}}]{Ramananarivo19}%
  \BibitemOpen
  \bibfield  {author} {\bibinfo {author} {\bibfnamefont {S.}~\bibnamefont
  {Ramananarivo}}, \bibinfo {author} {\bibfnamefont {E.}~\bibnamefont
  {Ducrot}}, \ and\ \bibinfo {author} {\bibfnamefont {J.}~\bibnamefont
  {Palacci}},\ }\bibfield  {title} {\enquote {\bibinfo {title}
  {Activity-controlled annealing of colloidal monolayers},}\ }\href {\doibase
  10.1038/s41467-019-11362-y} {\bibfield  {journal} {\bibinfo  {journal}
  {Nature Commun.}\ }\textbf {\bibinfo {volume} {10}},\ \bibinfo {pages} {3380}
  (\bibinfo {year} {2019})}\BibitemShut {NoStop}%
\bibitem [{\citenamefont {Chepizhko}\ \emph {et~al.}(2013)\citenamefont
  {Chepizhko}, \citenamefont {Altmann},\ and\ \citenamefont
  {Peruani}}]{Chepizhko13}%
  \BibitemOpen
  \bibfield  {author} {\bibinfo {author} {\bibfnamefont {O.}~\bibnamefont
  {Chepizhko}}, \bibinfo {author} {\bibfnamefont {E.~G.}\ \bibnamefont
  {Altmann}}, \ and\ \bibinfo {author} {\bibfnamefont {F.}~\bibnamefont
  {Peruani}},\ }\bibfield  {title} {\enquote {\bibinfo {title} {Optimal noise
  maximizes collective motion in heterogeneous media},}\ }\href {\doibase
  10.1103/PhysRevLett.110.238101} {\bibfield  {journal} {\bibinfo  {journal}
  {Phys. Rev. Lett.}\ }\textbf {\bibinfo {volume} {110}},\ \bibinfo {pages}
  {238101} (\bibinfo {year} {2013})}\BibitemShut {NoStop}%
\bibitem [{\citenamefont {Hinrichsen}(2000)}]{Hinrichsen00}%
  \BibitemOpen
  \bibfield  {author} {\bibinfo {author} {\bibfnamefont {H.}~\bibnamefont
  {Hinrichsen}},\ }\bibfield  {title} {\enquote {\bibinfo {title}
  {Non-equilibrium critical phenomena and phase transitions into absorbing
  states},}\ }\href {\doibase 10.1080/00018730050198152} {\bibfield  {journal}
  {\bibinfo  {journal} {Adv. Phys.}\ }\textbf {\bibinfo {volume} {49}},\
  \bibinfo {pages} {815--958} (\bibinfo {year} {2000})}\BibitemShut {NoStop}%
\bibitem [{\citenamefont {Corte}\ \emph {et~al.}(2008)\citenamefont {Corte},
  \citenamefont {Chaikin}, \citenamefont {Gollub},\ and\ \citenamefont
  {Pine}}]{Corte08}%
  \BibitemOpen
  \bibfield  {author} {\bibinfo {author} {\bibfnamefont {L.}~\bibnamefont
  {Corte}}, \bibinfo {author} {\bibfnamefont {P.~M.}\ \bibnamefont {Chaikin}},
  \bibinfo {author} {\bibfnamefont {J.~P.}\ \bibnamefont {Gollub}}, \ and\
  \bibinfo {author} {\bibfnamefont {D.~J.}\ \bibnamefont {Pine}},\ }\bibfield
  {title} {\enquote {\bibinfo {title} {Random organization in periodically
  driven systems},}\ }\href {\doibase 10.1038/nphys891} {\bibfield  {journal}
  {\bibinfo  {journal} {Nature Phys.}\ }\textbf {\bibinfo {volume} {4}},\
  \bibinfo {pages} {420--424} (\bibinfo {year} {2008})}\BibitemShut {NoStop}%
\bibitem [{\citenamefont {Reichhardt}\ and\ \citenamefont
  {Reichhardt}(2014{\natexlab{b}})}]{Reichhardt14a}%
  \BibitemOpen
  \bibfield  {author} {\bibinfo {author} {\bibfnamefont {C.}~\bibnamefont
  {Reichhardt}}\ and\ \bibinfo {author} {\bibfnamefont {C.~J.~Olson}\
  \bibnamefont {Reichhardt}},\ }\bibfield  {title} {\enquote {\bibinfo {title}
  {Absorbing phase transitions and dynamic freezing in running active matter
  systems},}\ }\href {\doibase 10.1039/c4sm01273a} {\bibfield  {journal}
  {\bibinfo  {journal} {Soft Matter}\ }\textbf {\bibinfo {volume} {10}},\
  \bibinfo {pages} {7502--7510} (\bibinfo {year}
  {2014}{\natexlab{b}})}\BibitemShut {NoStop}%
\bibitem [{\citenamefont {Wan}\ \emph {et~al.}(2008)\citenamefont {Wan},
  \citenamefont {Olson~Reichhardt}, \citenamefont {Nussinov},\ and\
  \citenamefont {Reichhardt}}]{Wan08}%
  \BibitemOpen
  \bibfield  {author} {\bibinfo {author} {\bibfnamefont {M.~B.}\ \bibnamefont
  {Wan}}, \bibinfo {author} {\bibfnamefont {C.~J.}\ \bibnamefont
  {Olson~Reichhardt}}, \bibinfo {author} {\bibfnamefont {Z.}~\bibnamefont
  {Nussinov}}, \ and\ \bibinfo {author} {\bibfnamefont {C.}~\bibnamefont
  {Reichhardt}},\ }\bibfield  {title} {\enquote {\bibinfo {title}
  {Rectification of swimming bacteria and self-driven particle systems by
  arrays of asymmetric barriers},}\ }\href {\doibase
  10.1103/PhysRevLett.101.018102} {\bibfield  {journal} {\bibinfo  {journal}
  {Phys. Rev. Lett.}\ }\textbf {\bibinfo {volume} {101}},\ \bibinfo {pages}
  {018102} (\bibinfo {year} {2008})}\BibitemShut {NoStop}%
\bibitem [{\citenamefont {Reichhardt}\ and\ \citenamefont
  {Reichhardt}(2017{\natexlab{b}})}]{Reichhardt17a}%
  \BibitemOpen
  \bibfield  {author} {\bibinfo {author} {\bibfnamefont {C.~J.~Olson}\
  \bibnamefont {Reichhardt}}\ and\ \bibinfo {author} {\bibfnamefont
  {C.}~\bibnamefont {Reichhardt}},\ }\bibfield  {title} {\enquote {\bibinfo
  {title} {Ratchet effects in active matter systems},}\ }\href {\doibase
  10.1146/annurev-conmatphys-031016-025522} {\bibfield  {journal} {\bibinfo
  {journal} {Ann. Rev. Condens. Matter Phys.}\ }\textbf {\bibinfo {volume}
  {8}},\ \bibinfo {pages} {51--75} (\bibinfo {year}
  {2017}{\natexlab{b}})}\BibitemShut {NoStop}%
\bibitem [{\citenamefont {Borba}\ \emph {et~al.}(2020)\citenamefont {Borba},
  \citenamefont {Domingos}, \citenamefont {Moraes}, \citenamefont {Potiguar},\
  and\ \citenamefont {Ferreira}}]{Borba20}%
  \BibitemOpen
  \bibfield  {author} {\bibinfo {author} {\bibfnamefont {A.~D.}\ \bibnamefont
  {Borba}}, \bibinfo {author} {\bibfnamefont {Jorge L.~C.}\ \bibnamefont
  {Domingos}}, \bibinfo {author} {\bibfnamefont {E.~C.~B.}\ \bibnamefont
  {Moraes}}, \bibinfo {author} {\bibfnamefont {F.~Q.}\ \bibnamefont
  {Potiguar}}, \ and\ \bibinfo {author} {\bibfnamefont {W.~P.}\ \bibnamefont
  {Ferreira}},\ }\bibfield  {title} {\enquote {\bibinfo {title} {Controlling
  the transport of active matter in disordered lattices of asymmetrical
  obstacles},}\ }\href {\doibase 10.1103/PhysRevE.101.022601} {\bibfield
  {journal} {\bibinfo  {journal} {Phys. Rev. E}\ }\textbf {\bibinfo {volume}
  {101}},\ \bibinfo {pages} {022601} (\bibinfo {year} {2020})}\BibitemShut
  {NoStop}%
\bibitem [{\citenamefont {O'Byrne}\ \emph {et~al.}(2022)\citenamefont
  {O'Byrne}, \citenamefont {Kafri}, \citenamefont {Tailleur},\ and\
  \citenamefont {van Wijland}}]{Obyrne22}%
  \BibitemOpen
  \bibfield  {author} {\bibinfo {author} {\bibfnamefont {J.}~\bibnamefont
  {O'Byrne}}, \bibinfo {author} {\bibfnamefont {Y.}~\bibnamefont {Kafri}},
  \bibinfo {author} {\bibfnamefont {J.}~\bibnamefont {Tailleur}}, \ and\
  \bibinfo {author} {\bibfnamefont {F.}~\bibnamefont {van Wijland}},\
  }\bibfield  {title} {\enquote {\bibinfo {title} {Time irreversibility in
  active matter, from micro to macro},}\ }\href {\doibase
  10.1038/s42254-021-00406-2} {\bibfield  {journal} {\bibinfo  {journal}
  {Nature Rev. Phys.}\ }\textbf {\bibinfo {volume} {4}},\ \bibinfo {pages}
  {167--183} (\bibinfo {year} {2022})}\BibitemShut {NoStop}%
\bibitem [{\citenamefont {Weber}\ \emph {et~al.}(2013)\citenamefont {Weber},
  \citenamefont {Hauschild}, \citenamefont {Schwarz}, \citenamefont {Moussion},
  \citenamefont {de~Vries}, \citenamefont {Legler}, \citenamefont {Luther},
  \citenamefont {Bollenbach},\ and\ \citenamefont {Sixt}}]{Weber13}%
  \BibitemOpen
  \bibfield  {author} {\bibinfo {author} {\bibfnamefont {M.}~\bibnamefont
  {Weber}}, \bibinfo {author} {\bibfnamefont {R.}~\bibnamefont {Hauschild}},
  \bibinfo {author} {\bibfnamefont {J.}~\bibnamefont {Schwarz}}, \bibinfo
  {author} {\bibfnamefont {C.}~\bibnamefont {Moussion}}, \bibinfo {author}
  {\bibfnamefont {I.}~\bibnamefont {de~Vries}}, \bibinfo {author}
  {\bibfnamefont {D.~F.}\ \bibnamefont {Legler}}, \bibinfo {author}
  {\bibfnamefont {S.~A.}\ \bibnamefont {Luther}}, \bibinfo {author}
  {\bibfnamefont {T.}~\bibnamefont {Bollenbach}}, \ and\ \bibinfo {author}
  {\bibfnamefont {M.}~\bibnamefont {Sixt}},\ }\bibfield  {title} {\enquote
  {\bibinfo {title} {Interstitial dendritic cell guidance by haptotactic
  chemokine gradients},}\ }\href {\doibase 10.1126/science.1228456} {\bibfield
  {journal} {\bibinfo  {journal} {Science}\ }\textbf {\bibinfo {volume}
  {339}},\ \bibinfo {pages} {328--332} (\bibinfo {year} {2013})}\BibitemShut
  {NoStop}%
\bibitem [{\citenamefont {Tweedy}\ \emph {et~al.}(2016)\citenamefont {Tweedy},
  \citenamefont {Susanto},\ and\ \citenamefont {Insall}}]{Tweedy16}%
  \BibitemOpen
  \bibfield  {author} {\bibinfo {author} {\bibfnamefont {L.}~\bibnamefont
  {Tweedy}}, \bibinfo {author} {\bibfnamefont {O.}~\bibnamefont {Susanto}}, \
  and\ \bibinfo {author} {\bibfnamefont {R.~H.}\ \bibnamefont {Insall}},\
  }\bibfield  {title} {\enquote {\bibinfo {title} {Self-generated chemotactic
  gradients - cells steering themselves},}\ }\href {\doibase
  10.1016/j.ceb.2016.04.003} {\bibfield  {journal} {\bibinfo  {journal} {Curr.
  Opin. Cell Biol.}\ }\textbf {\bibinfo {volume} {42}},\ \bibinfo {pages}
  {46--51} (\bibinfo {year} {2016})}\BibitemShut {NoStop}%
\bibitem [{\citenamefont {Don{\' a}}\ \emph {et~al.}(2013)\citenamefont {Don{\'
  a}}, \citenamefont {Barry}, \citenamefont {Valentin}, \citenamefont {Quirin},
  \citenamefont {Khmelinskii}, \citenamefont {Kunze}, \citenamefont {Durdu},
  \citenamefont {Newton}, \citenamefont {Fernandez-Minan}, \citenamefont
  {Huber}, \citenamefont {Knop},\ and\ \citenamefont {Gilmour}}]{Dona13}%
  \BibitemOpen
  \bibfield  {author} {\bibinfo {author} {\bibfnamefont {E.}~\bibnamefont
  {Don{\' a}}}, \bibinfo {author} {\bibfnamefont {J.~D.}\ \bibnamefont
  {Barry}}, \bibinfo {author} {\bibfnamefont {G.}~\bibnamefont {Valentin}},
  \bibinfo {author} {\bibfnamefont {C.}~\bibnamefont {Quirin}}, \bibinfo
  {author} {\bibfnamefont {A.}~\bibnamefont {Khmelinskii}}, \bibinfo {author}
  {\bibfnamefont {A.}~\bibnamefont {Kunze}}, \bibinfo {author} {\bibfnamefont
  {S.}~\bibnamefont {Durdu}}, \bibinfo {author} {\bibfnamefont {L.~R.}\
  \bibnamefont {Newton}}, \bibinfo {author} {\bibfnamefont {A.}~\bibnamefont
  {Fernandez-Minan}}, \bibinfo {author} {\bibfnamefont {W.}~\bibnamefont
  {Huber}}, \bibinfo {author} {\bibfnamefont {M.}~\bibnamefont {Knop}}, \ and\
  \bibinfo {author} {\bibfnamefont {D.}~\bibnamefont {Gilmour}},\ }\bibfield
  {title} {\enquote {\bibinfo {title} {Directional tissue migration through a
  self-generated chemokine gradient},}\ }\href {\doibase 10.1038/nature12635}
  {\bibfield  {journal} {\bibinfo  {journal} {Nature (London)}\ }\textbf
  {\bibinfo {volume} {503}},\ \bibinfo {pages} {285--289} (\bibinfo {year}
  {2013})}\BibitemShut {NoStop}%
\bibitem [{\citenamefont {Solon}\ \emph {et~al.}(2006)\citenamefont {Solon},
  \citenamefont {Streicher}, \citenamefont {Richter}, \citenamefont
  {Brochard-Wyart},\ and\ \citenamefont {Bassereau}}]{Solon06}%
  \BibitemOpen
  \bibfield  {author} {\bibinfo {author} {\bibfnamefont {J.}~\bibnamefont
  {Solon}}, \bibinfo {author} {\bibfnamefont {P.}~\bibnamefont {Streicher}},
  \bibinfo {author} {\bibfnamefont {R.}~\bibnamefont {Richter}}, \bibinfo
  {author} {\bibfnamefont {F.}~\bibnamefont {Brochard-Wyart}}, \ and\ \bibinfo
  {author} {\bibfnamefont {P.}~\bibnamefont {Bassereau}},\ }\bibfield  {title}
  {\enquote {\bibinfo {title} {Vesicles surfing on a lipid bilayer:
  self-induced haptotactic motion},}\ }\href {\doibase 10.1073/pnas.0601400103}
  {\bibfield  {journal} {\bibinfo  {journal} {Proc. Natl. Acad. Sci. (USA)}\
  }\textbf {\bibinfo {volume} {103}},\ \bibinfo {pages} {12382} (\bibinfo
  {year} {2006})}\BibitemShut {NoStop}%
\bibitem [{\citenamefont {Cochet-Escartin}\ \emph {et~al.}(2021)\citenamefont
  {Cochet-Escartin}, \citenamefont {Demircigil}, \citenamefont {Hirose},
  \citenamefont {Allais}, \citenamefont {Gonzalo}, \citenamefont {Mikaelian},
  \citenamefont {Funamoto}, \citenamefont {Anjard}, \citenamefont {Calvez},\
  and\ \citenamefont {Rieu}}]{CochetEscartin21}%
  \BibitemOpen
  \bibfield  {author} {\bibinfo {author} {\bibfnamefont {O.}~\bibnamefont
  {Cochet-Escartin}}, \bibinfo {author} {\bibfnamefont {M.}~\bibnamefont
  {Demircigil}}, \bibinfo {author} {\bibfnamefont {S.}~\bibnamefont {Hirose}},
  \bibinfo {author} {\bibfnamefont {B.}~\bibnamefont {Allais}}, \bibinfo
  {author} {\bibfnamefont {P.}~\bibnamefont {Gonzalo}}, \bibinfo {author}
  {\bibfnamefont {I.}~\bibnamefont {Mikaelian}}, \bibinfo {author}
  {\bibfnamefont {K.}~\bibnamefont {Funamoto}}, \bibinfo {author}
  {\bibfnamefont {C.}~\bibnamefont {Anjard}}, \bibinfo {author} {\bibfnamefont
  {V.}~\bibnamefont {Calvez}}, \ and\ \bibinfo {author} {\bibfnamefont {J.-P.}\
  \bibnamefont {Rieu}},\ }\bibfield  {title} {\enquote {\bibinfo {title}
  {Hypoxia triggers collective aerotactic migration in {\it dictyostelium
  discoideum}},}\ }\href {\doibase 10.7554/eLife.64731} {\bibfield  {journal}
  {\bibinfo  {journal} {eLife}\ }\textbf {\bibinfo {volume} {10}},\ \bibinfo
  {pages} {e64731} (\bibinfo {year} {2021})}\BibitemShut {NoStop}%
\bibitem [{\citenamefont {Sengupta}\ \emph {et~al.}(2009)\citenamefont
  {Sengupta}, \citenamefont {van Teeffelen},\ and\ \citenamefont {L{\"
  o}wen}}]{Sengupta09}%
  \BibitemOpen
  \bibfield  {author} {\bibinfo {author} {\bibfnamefont {A.}~\bibnamefont
  {Sengupta}}, \bibinfo {author} {\bibfnamefont {S.}~\bibnamefont {van
  Teeffelen}}, \ and\ \bibinfo {author} {\bibfnamefont {H.}~\bibnamefont {L{\"
  o}wen}},\ }\bibfield  {title} {\enquote {\bibinfo {title} {Dynamics of a
  microorganism moving by chemotaxis in its own secretion},}\ }\href {\doibase
  10.1103/PhysRevE.80.031122} {\bibfield  {journal} {\bibinfo  {journal} {Phys.
  Rev. E}\ }\textbf {\bibinfo {volume} {80}},\ \bibinfo {pages} {031122}
  (\bibinfo {year} {2009})}\BibitemShut {NoStop}%
\bibitem [{\citenamefont {Liebchen}\ and\ \citenamefont {L{\"
  o}wen}(2018)}]{Liebchen18}%
  \BibitemOpen
  \bibfield  {author} {\bibinfo {author} {\bibfnamefont {B.}~\bibnamefont
  {Liebchen}}\ and\ \bibinfo {author} {\bibfnamefont {H.}~\bibnamefont {L{\"
  o}wen}},\ }\bibfield  {title} {\enquote {\bibinfo {title} {Synthetic
  chemotaxis and collective behavior in active matter},}\ }\href {\doibase
  10.1021/acs.accounts.8b00215} {\bibfield  {journal} {\bibinfo  {journal}
  {Acc. Chem. Res.}\ }\textbf {\bibinfo {volume} {51}},\ \bibinfo {pages}
  {2982--2990} (\bibinfo {year} {2018})}\BibitemShut {NoStop}%
\bibitem [{\citenamefont {Lavergne}\ \emph {et~al.}(2019)\citenamefont
  {Lavergne}, \citenamefont {Wendehenne}, \citenamefont {Baeuerle},\ and\
  \citenamefont {Bechinger}}]{Lavergne19}%
  \BibitemOpen
  \bibfield  {author} {\bibinfo {author} {\bibfnamefont {F.~A.}\ \bibnamefont
  {Lavergne}}, \bibinfo {author} {\bibfnamefont {H.}~\bibnamefont
  {Wendehenne}}, \bibinfo {author} {\bibfnamefont {T.}~\bibnamefont
  {Baeuerle}}, \ and\ \bibinfo {author} {\bibfnamefont {C.}~\bibnamefont
  {Bechinger}},\ }\bibfield  {title} {\enquote {\bibinfo {title} {Group
  formation and cohesion of active particles with visual perception-dependent
  motility},}\ }\href {\doibase 10.1126/science.aau5347} {\bibfield  {journal}
  {\bibinfo  {journal} {Science}\ }\textbf {\bibinfo {volume} {364}},\ \bibinfo
  {pages} {70} (\bibinfo {year} {2019})}\BibitemShut {NoStop}%
\bibitem [{\citenamefont {Yang}\ \emph {et~al.}(2021)\citenamefont {Yang},
  \citenamefont {Huang}, \citenamefont {Zhao},\ and\ \citenamefont
  {Zhang}}]{Yang21}%
  \BibitemOpen
  \bibfield  {author} {\bibinfo {author} {\bibfnamefont {S.}~\bibnamefont
  {Yang}}, \bibinfo {author} {\bibfnamefont {M.}~\bibnamefont {Huang}},
  \bibinfo {author} {\bibfnamefont {Y.}~\bibnamefont {Zhao}}, \ and\ \bibinfo
  {author} {\bibfnamefont {H.~P.}\ \bibnamefont {Zhang}},\ }\bibfield  {title}
  {\enquote {\bibinfo {title} {Controlling cell motion and microscale flow with
  polarized light fields},}\ }\href {\doibase 10.1103/PhysRevLett.126.058001}
  {\bibfield  {journal} {\bibinfo  {journal} {Phys. Rev. Lett.}\ }\textbf
  {\bibinfo {volume} {126}},\ \bibinfo {pages} {058001} (\bibinfo {year}
  {2021})}\BibitemShut {NoStop}%
\bibitem [{\citenamefont {Bi}\ \emph {et~al.}(2016)\citenamefont {Bi},
  \citenamefont {Yang}, \citenamefont {Marchetti},\ and\ \citenamefont
  {Manning}}]{Bi16}%
  \BibitemOpen
  \bibfield  {author} {\bibinfo {author} {\bibfnamefont {D.}~\bibnamefont
  {Bi}}, \bibinfo {author} {\bibfnamefont {X.}~\bibnamefont {Yang}}, \bibinfo
  {author} {\bibfnamefont {M.~C.}\ \bibnamefont {Marchetti}}, \ and\ \bibinfo
  {author} {\bibfnamefont {M.~L.}\ \bibnamefont {Manning}},\ }\bibfield
  {title} {\enquote {\bibinfo {title} {Motility-driven glass and jamming
  transitions in biological tissues},}\ }\href {\doibase
  10.1103/PhysRevX.6.021011} {\bibfield  {journal} {\bibinfo  {journal} {Phys.
  Rev. X}\ }\textbf {\bibinfo {volume} {6}},\ \bibinfo {pages} {021011}
  (\bibinfo {year} {2016})}\BibitemShut {NoStop}%
\end{thebibliography}%
\end{document}